Correlated Quantum Phenomena of Spin-Orbit Coupled Perovskite Oxide Heterostructures: Cases of $SrRuO_3$ and $SrIrO_3$-Based Artificial Superlattices

*Seung Gyo Jeong, Jin Young Oh, Lin Hao*, Jian Liu*, and Woo Seok Choi**

S. G. Jeong, J. Y. Oh, W. S. Choi
Department of Physics, Sungkyunkwan University, Suwon 16419, Korea
E-mail: choiws@skku.edu

Lin Hao
Anhui Key Laboratory of Condensed Matter Physics at Extreme Conditions, High Magnetic Field Laboratory, HFIPS, Chinese Academy of Sciences, Hefei, Anhui 230031, China
E-mail: haolin@hmfl.ac.cn

Jian Liu
Department of Physics and Astronomy, University of Tennessee, Knoxville, Tennessee 37996, USA
E-mail: jianliu@utk.edu

Unexpected, yet useful functionalities emerge when two or more materials merge coherently. Artificial oxide superlattices realize atomic and crystal structures that are not available in nature, thus providing controllable correlated quantum phenomena. This review focuses on 4*d* and 5*d* oxide superlattices, in which the spin-orbit coupling plays a significant role compared with conventional 3*d* oxide superlattices. Modulations in crystal structures with octahedral distortion, phonon engineering, electronic structures, spin orderings, and dimensionality control are discussed for 4*d* oxide superlattices. Atomic and magnetic structures, $J_{\mathrm{eff}} = 1/2$ pseudospin and charge fluctuations, and the integration of topology and correlation are discussed for 5*d* oxide superlattices. This review provides insights into how correlated quantum phenomena arise from the deliberate design of superlattice structures that give birth to novel functionalities.

# 1. Introduction

## 1.1 Oxide Heterostructures

Heterostructures composed of transition metal oxides have been celebrated for decades in fundamental condensed matter research as they serve as one of the key material systems for the development of future functional devices, including thermal, optical, electronic, magnetic, and energy devices.[1-8] Because of the strong internal electric field generated by the oxygen ions, the chemical bonding nature between the various transition metal elements and oxygen ions are either covalent or ionic.[9] This strong chemical bonding ensures highly coherent lattice matching epitaxial relationship between the thin films and the substrates or between the different layers within oxide heterostructures.[4, 10] Such coherent structure is not attainable in structures with, e.g., van der Waals bonding.[11, 12]This feature also distinguishes oxide heterostructures from semiconductor heterostructures, which shape the contemporary electronics industry, by embedding correlated functionalities for further scalable and energy-efficient multifunctional devices. Hence, the idea of stacking the building blocks of transition metal oxides with various physical and chemical properties to exploit synergetic functionality has been prevalent in the field. Oxide heterostructures can ideally realize combinatorial devices that can simultaneously sense, compute, memorize, and actuate by epitaxially combining the necessary layers. Moreover, the functionalities of each layer can be altered as they are being combined, thereby leading to emergent synergetic functionalities. Thus, the opportunity to develop new materials has been maximized.

Epitaxial oxide heterostructures have led to a revolutionary expansion in the fundamental physical and chemical sciences.[4] In addition to the conventional strong correlation among charge, spin, lattice, and orbital degrees of freedom in a single transition metal oxide compound, oxide heterostructures of transition metal oxides provide important tuning parameters,[5] such as epitaxial strain, octahedral distortion, structural symmetry constraints, charge transfer, magnetic and superconducting proximity, orbital engineering, interfacial band bending, ferroelectric field effect, quantum tunneling, dimensional crossover, and inversion symmetry breaking. The systematic and facile control of these parameters has deepened our understanding of the underlying mechanisms of the functionalities.[13] In particular, the on-demand growth of oxide heterostructures has allowed us to explore the vast landscape of intriguing and important problems in the field.

Oxide superlattices employ coherent lattice structures of the oxide heterostructures to the very limit, which is a single atomic layer.[4] They promote single-crystalline structures that are not available in nature, thereby overcoming the thermodynamic constraints for spontaneous natural crystal formation. The development of atomic-scale precision control of the layers enabled genuine man-made crystals with facile tunability of the lattice structures and periodicity. Even though the strong correlation often occurring in oxides largely modify the structure, the electronic band structure is essentially the energy eigenfunction of electrons under a periodic potential. Hence, the control of the periodic potential via artificial engineering of lattices fundamentally allows altering the physical and chemical properties of the material. For example, the size of the unit cell of superlattices can be digitally increased with the periodicity superimposing the original periodicity of each constituent layer, giving rise to the band folding in the momentum space with the formation of minibands. By exploiting the aforementioned strong covalent/ionic bonding between the transition metal and oxygen, the crystalline symmetry, chemical stoichiometry, and dimensionality of the artificial crystal can be further modified. The strong covalent/ionic bonding also underscores the importance of the functional interfaces, the effect of which is considerably amplified in superlattices.

The numerous oxide superlattices can be classified in several ways. For example, tri-color superlattices, in which the inversion symmetry is broken, can be distinguished from conventional bi-color superlattices.[10] Additionally, the global structures of the superlattices, for example, perovskite, spinel, pyrochlore, and brownmillerite, can be a basis for classification. However, the vast majority of oxide superlattices studied to date have perovskite structures. One can also categorize the oxide superlattices by focusing on the transition metal elements. Whereas 3*d* transition metal oxides have been intensively studied over the last several decades, recently, 4*d* and 5*d* transition metal oxides have become increasingly important with the introduction of appreciable spin-orbit coupling ( $\lambda$ ), of which the strength is comparable to the on-site Coulomb interaction ($U$) and crystal field splitting energy ($\Delta$) within the transition metal oxygen octahedra (Table1). As a result of the variety of spin-orbit-entangled degrees of freedom and their non-trivial interactions, various exotic correlated phases can realize in 4*d* and 5*d* elements, including spin-orbit-assisted Mott insulators, $J_{\mathrm{eff}} = 1/2$ magnetism, quantum spin liquids, excitonic magnetism, multipolar orderings, and correlated topological semimetals.[14] The combination of these novel spin-orbit-entangled states with conventional oxide superlattice approaches provides a new frontier of correlated quantum phenomena with practical controllability that inspires future quantum electronic devices with various functionalities. Therefore, oxide superlattices containing 4*d* and 5*d* transition metal oxide layers possess physics beyond in those superlattices with 3*d* transition metal elements, as schematically shown by the correlated functionalities in **Figure 1**. Herein, first, we briefly summarize previous studies on 3*d* oxide superlattices, the idea and methods of which can be applied and extended to 4*d* and 5*d* oxide superlattices. Subsequently, we discuss the 4*d* and 5*d* oxide superlattices in further depth.

### 1.2 3*d* Based Superlattice

Most traditional 3*d* perovskite oxides have been used as constituent layers in superlattices. Titanates ($SrTiO_3$, $BaTiO_3$, $CaTiO_3$, and $LaTiO_3$) have been extensively studied, partly owing to the intriguing physical behavior of $SrTiO_3$, including its quantum paraelectric nature and ability to accommodate oxygen vacancies. Before the discovery of unexpected metallic behavior at the interface between two band insulators, i.e., $LaAlO_3$ and $SrTiO_3$,[15] the interface between a Mott insulator and a band insulator, i.e., $LaTiO_3$ and $SrTiO_3$, respectively, was found to be metallic owing to the "electronic reconstruction".[16] $LaTiO_3/SrTiO_3$ superlattices provided an ideal testbed for experimental examination of the unexpected metallic interface as they effectively amplify the conducting signature observed not only from the transport measurements but also by optical spectroscopy. As shown in **Figure 2a**, the structure can be viewed as a La substitution for Sr within the $SrTiO_3$ matrix in a layer-by-layer manner, i.e., delta-doping. Optical spectroscopy characterizes the intrinsic carrier density of the entire superlattice, as shown in **Figure 2b**.[17] Here, the sheet carrier density is independent of the overall La concentration, thus indicating that only the interface is metallic and not the entire sample. Note that such observation would have been extremely difficult with a single interface in conventional oxide heterostructures. The $LaTiO_3/SrTiO_3$ interface has the potential for use in transparent conducting electrode applications as the visible light can penetrate the sample without much absorption.[18] The La delta-doping approach was later developed as fractional doping of the superlattice, thus enabling control of the carrier density at the interfaces.[19] This approach was found to be beneficial for thermoelectric applications.[20] Another important functionality of $LaTiO_3/SrTiO_3$ superlattices is quantum resonant tunneling (**Figure 2c**).[21] This report is the first observation of resonant tunneling in oxide superlattices, which has been an important subject in semiconductor superlattices. On the other hand, the dimensional quantum confinement of $3d^1$ electrons was optically characterized in the $LaTiO_3/LaAlO_3$ superlattices.[22]

As in the case of La doping, $SrTiO_3$ is a very effective system that exhibits a wide variety of electronic phases (from insulator to superconductor) upon doping. Nb doping has proven to be efficient, and by constructing superlattices of metallic Nb-doped $SrTiO_3$ and $SrTiO_3$ layers, the dimensionality of the metallic layer can be reduced to the thickness of a single unit-cell. One of the merits of the reduced dimension is the enhancement of the electron-phonon coupling, which increases the thermopower of the low-dimensional system, as shown in **Figure 2d**.[23, 24]

Titanate superlattices have been extensively studied in terms of their ferroelectric properties. $BaTiO_3$ is ferroelectric, $SrTiO_3$ is incipient ferroelectric, and $CaTiO_3$ is paraelectric. However, by combining these materials at the atomic scale, the ferroelectricity can be further enhanced, as shown in **Figure 2e**. The paraelectric layers become ferroelectric owing to cooperative structural distortion within the superlattices.[25] On the other hand, the three titanate perovskites can be combined to form tri-color superlattices with inherent inversion symmetry breaking. Inversion symmetry breaking is essential for the emergence of ferroelectricity, and ferroelectricity may be expected even with nonferroelectric constituents. The tri-color superlattices exhibit enhanced ferroelectric polarization, thus corroborating the intuitive symmetry-breaking mechanism (**Figure 2f**).[10] Such a polar order could be further modified by geometric constraints offered by the oxide heterostructure. The epitaxial tensile strain in $PbTiO_3$/$SrTiO_3$ multilayers induced flux-closures,[26] which was later extended to smaller polar vortex structures in $PbTiO_3$/$SrTiO_3$ superlattices, as shown in **Figure 2g**.[27] Furthermore, the modulation of the lattice structure and interfaces leads to an entirely different global vibrational response of the superlattices (**Figure 2h**), thereby leading to a greater control over the thermal conductivity.[28, 29]

Except for titanates, 3*d* perovskite superlattices have been realized using vanadates,[30] manganites,[31] ferrites,[32] cobaltites,[33] and nickelates.[34] In nickelates, the dimensionality change results in a collective change in the correlated electronic and magnetic phase transitions (**Figure 2i**). For oxide superlattices other than perovskites, hexagonal $LuFeO_3$ superlattices have representatively demonstrated magnetoelectric coupling promoted by geometric ferroelectricity (**Figure 2j**).[35]

This review summarizes recent studies on 4*d* and 5*d* oxide superlattices. Most of the 4*d* and 5*d* superlattices studied thus far contain $SrRuO_3$ and $SrIrO_3$ layers, respectively. Hence, herein, we focus on the superlattices of these two important materials. Chapter 2 introduces state-of-the-art atomic-scale epitaxy techniques that enable the realization of artificial superlattices. Chapter 3 discusses the $SrRuO_3$ based superlattices. Starting from the lattice structure modulation, we discuss artificial phonon engineering, ferromagnetism, dimensionality-induced phase transition, and topologically nontrivial spin structures. In Chapter 4, we discuss $SrIrO_3$ based superlattices. The crystal structure, epitaxial strain, spin-charge fluctuation, and fluctuation of $J_{\mathrm{eff}} = 1/2$ pseudospin in the 2D limit are discussed. We will further compare the natural layered crystal of the Ruddlesden-Popper series with the superlattices. Finally, we summarize this review in Chapter 5 with an outlook.

## 2. Atomic Scale Epitaxy Techniques: Realization of Artificial Oxide Crystal

The development of modern epitaxy techniques has provided atomic-scale precision controllability of state-of-the-art oxide heterostructures. Prominent epitaxy techniques for realizing oxide superlattices include sputter deposition, pulsed laser epitaxy, and molecular beam epitaxy. Sputtering was first employed to fabricate iron oxide films in 1852.[36, 37] To this day, sputtering is widely used for thin film growth as it is economical and useful for mass production in the industry.[38]The sputtering process uses the plasma produced by electrical discharge and the electrostatic acceleration of ions to deposit the thin film material on the substrate surfaces. Oxide heterostructures and superlattices can be grown using two target compounds. High-quality oxide superlattices grown by sputter deposition include cuprates,[39] titanates,[38, 40] ruthenates,[41] nickelates,[42, 43] and manganites.[44]

Owing to the improvement in vacuum techniques in the 1960s, both pulsed laser epitaxy and molecular beam epitaxy have been developed.[45, 46] Pulsed laser epitaxy is one of the most appreciated approaches that demonstrates the atomic-scale epitaxy of oxide superlattices. The thin film growth process is as follows.[47] 1) A pulsed laser ablates the surface of a target in a vacuum chamber and generates a plasma plume. 2) The plasma plume propagates and reaches the heated substrate. 3) The ion clusters crystallize on the substrate surface through a thermodynamic process. Pulsed laser epitaxy is useful for synthesizing complex heterostructures and superlattices using a target manipulator. The growth window of pulsed laser epitaxy is typically wide for transition metal oxides and can be modulated using several experimental parameters, such as the laser (pulse energy and repetition rate), substrate (material, surface, and temperature), and type of background gas, pressure, and flow rate in the growth chamber.[48] In particular, the interaction (or scattering) between plasma and background determines the kinetics, oxidation state, and stoichiometry of the incoming plume, as well as the final surface dynamics of incoming species. These factors settle the growth mechanisms of oxide thin films, such as layer-by-layer, step flow, and island growth modes.[47, 49-51] For example, in cases of $SrRuO_3$ based superlattices, the growth of the thin film layer of $SrRuO_3$ depends sensitively on the substrate temperature, oxygen background pressure, and laser fluence. First, due to the volatile nature of Ru ions at high temperatures, substrate temperatures of around 700 °C are commonly used. Second, a sufficiently oxygen-rich environment of 100 mTorr is required to synthesize stoichiometric $SrRuO_3$ layers.[49, 52] Third, since high laser fluence can induce unintended cation vacancies and change the growth modes,[53, 54] a laser fluence of around 1.2 $J/cm^2$ is typically used for the growth of $SrRuO_3$ layers. Finally, when constructing a superlattice, we should consider the deposition conditions of the partner layer, ensuring that they do not compromise the functionality of $SrRuO_3$. Moreover, it is crucial to experimentally verify whether the selected deposition conditions are appropriate. Although measuring small changes in the chemical composition ratio of $SrRuO_3$ according to these experimental conditions is difficult, the prominent observations are the variation in the ferromagnetic phase transition temperature and electrical transport behaviors.[49, 53] On the other hand, by adjusting the number of laser pulses, the growth of each layer within a superlattice can be accurately controlled with atomic-scale precision, even without reflection high energy electron diffraction.[55] Thanks to the versatile technical advantages, the majority of oxide superlattices with atomic-scale precision thickness control have been reported as using pulsed laser epitaxy.

Molecular beam epitaxy supplies each constituent element of perovskite oxide independently, under ultrahigh vacuum conditions, whereas both sputtering and pulsed laser epitaxy typically use stoichiometric ceramic targets. Molecular beam epitaxy offers selective cation control of perovskite oxides with atomic-scale precision thickness control via the instantaneous shutter control of individual elements.[56] High-vacuum conditions also provide a high mean free path for electrons and ions suitable for in-situ characterizations, including high-energy electron diffraction, low-energy electron diffraction, X-ray photoelectron spectroscopy, and Auger electron spectroscopy. One of the experimental parameters of oxide molecular beam epitaxy for stoichiometric films is the oxidant and/or the oxidant background pressure. Because molecular oxygen in early oxide molecular beam epitaxy is insufficient in numerous cases, oxygen plasma and/or ozone gases have become popular for achieving more aggressive oxidizing environments; however, they are accompanied by high vapor pressure and require external vapor inlet systems. More recently, solid source metal-organic molecular beam epitaxy has been suggested to obtain low vapor pressures and low source temperatures, which is more stable and cheaper compared with other oxide molecular beam epitaxies.[57] Finally, atomic layer deposition has also proven to be useful in depositing oxide thin films, although the realization of perovskite oxide superlattices has yet to be demonstrated.

## 3. 4*d* $SrRuO_3$ Based Superlattices

The 4*d* perovskite oxide $SrRuO_3$ exhibits itinerant ferromagnetism with various correlated functionalities.[58-60] In bulk, $SrRuO_3$ exhibits an orthorhombic structure at room temperature with the *Pbnm* space group having $a^-a^-c^+$ rotation pattern in the Glazer notation. The bulk lattice constants are $a$ = 5.5670 Å, $b$ = 5.5304 Å, and $c$ = 7.8446 Å in orthorhombic notation,[61] whereas the pseudo-cubic lattice constant is 3.93 Å.[49] Owing to their chemical stability, electrical conductivity, and versatile physical properties, $SrRuO_3$ films have received considerable attention for both fundamental correlated physics and applications.[58] In particular, the spin-orbit coupling strength of the 4*d* Ru ions is comparable to the electronic correlation energy scales, for example, Hund's coupling, crystal field splitting, and electron hopping,[14] thus leading to novel spin-orbit-coupled electronic ground states and physical phenomenon. The three Ru $t_{2g}$ orbitals ($d_{xy}$, $d_{xz}$, and $d_{yz}$) within $SrRuO_3$ are filled with four electrons following Hund's rule, thereby stabilizing $S = 1$ spin states in the bulk.[62] Theoretical calculations suggested a half-metallic behavior with Stoner spin splitting of $SrRuO_3$.[63] When the stoichiometry deviates from the ideal $SrRuO_3$, for example, when Ru-O vacancies are introduced, a structural phase transition from orthorhombic to tetragonal is observed, along with changes in the electronic and chemical behaviors.[49, 64]

By taking advantage of the epitaxial heterostructuring, facile control over the *d*-electron configuration, octahedral bonding geometry, crystalline symmetry, magnetic interactions, and electronic dimensionality can be achieved. Since the first growth of single-crystalline $SrRuO_3$ epitaxial films in 1992 reported by Eom et al.,[65] numerous $SrRuO_3$ heterostructures have been demonstrated to exhibit intriguing functional phenomena, such as metal-insulator transition,[66, 67] structural phase transition,[68, 69] electrocatalytic activity,[64] spin-phonon coupling,[70] itinerant ferromagnetism,[71, 72] controllable magnetic anisotropy,[52, 73] anomalous and/or topological Hall effect,[74-79] Weyl fermion behavior,[80] topological magnon,[81, 82] and chiral phonon-mediated magnetic interactions.[55] Several studies have demonstrated epitaxially grown $SrRuO_3$ thin films on $SiO_2$/Si substrates,[83-85] which implies the potential integration of correlated functional oxides with Si-based electronics. Atomically designed $SrRuO_3$ superlattices provide further extended controllability of the correlated functionalities, depending on the type of partner oxide layer and substrate, chemical stoichiometry, individual layer thickness, repetition number, and stacking order of the supercell structure. Table 2 summarizes the reported $SrRuO_3$ based superlattices with various functional partner compounds.[41, 54, 55, 70, 86-164] Herein, several emergent functionalities of $SrRuO_3$-based superlattices and their modulations are reviewed in the following Sections.

### 3.1 Modulation of Correlated Lattice Degree of Freedom

The customized lattice degree of freedom of orthorhombic $SrRuO_3$ provides a facile control knob for determining multiple correlated functionalities.[58] A theoretical study proposed that the orthorhombic structure of $SrRuO_3$ with octahedral tilt and the tetragonal structure of $SrRuO_3$ without octahedral tilt have almost the same energy, differing by only a few tens of meV.[165] By taking advantage of the nearly degenerate energy of the orthorhombic and tetragonal structures of $SrRuO_3$, it is possible to control the octahedral tilts of both structures and stabilize various structural phases through epitaxial heterostructures. Such structural modification modulate the ferromagnetic behavior (transition temperature, coercive field, magnetic easy axis, and magnetization value), electronic structure (crystal field and orbital hybridization), magnetotransports, and topological behavior. This section introduces various structural modulations in $SrRuO_3$ based superlattices from microscopic octahedral distortion to macroscopic crystalline symmetry.

#### *3.1.1 Octahedral Distortion*

Contemporary scanning transmission electron microscopy can directly visualize minute distortions of $RuO_6$ octahedra in $SrRuO_3$-based heterostructures. Wang et al. reported the $SrRuO_3$ thickness-dependent evolution of lattice mismatch within $SrRuO_3$/$LaCoO_3$ multilayers grown on $SrTiO_3$ substrates using selected-area electron diffraction patterns (**Figure 3a**).[156] The authors detected Ruddlesden-Popper-type faults induced by excessive Sr ions within the $SrRuO_3$ layers. The microstructure could be further modified by controlling the $SrRuO_3$ layer thickness. Jeong et al. demonstrated the propagation of octahedral distortions in $SrRuO_3$/$SrTiO_3$ superlattices. Even for the same thickness of the $SrRuO_3$layer, the octahedral distortion was systematically modulated by controlling the atomic-scale thickness of adjacent $SrTiO_3$ layers (**Figure 3b**).[91] As the reportedly, as the STO thickness increased, the cubic nature of the $SrTiO_3$ layer better restrained the octahedral distortion of the $SrRuO_3$ layer within the superlattices. A similar approach was adopted by Shan et al., in which the octahedral distortion in $SrRuO_3/SrCuO_2$ superlattices was modulated by dimensionality control of the infinite layered $SrCuO_2$ (**Figure 3c**).[117] With decreasing $SrCuO_2$ layer thickness, the $CuO_2$ infinite layer was shown to change from planar to chain-type, which modulated the polyhedral connectivity at the interface. On the other hand, Ziese et al. showed that the out-of-plane lattice constant for the $SrRuO_3$ layer in $SrRuO_3/Pr_{0.7}Ca_{0.3}MnO_3$ superlattices could be controlled by the thickness of the $Pr_{0.7}Ca_{0.3}MnO_3$ layer (**Figure 3d**).[141] Yao et al. suggested that the ferroelectric polarization of $BiFeO_3$ induced ionic displacements of the interfacial $SrRuO_3$ layer in the $SrRuO_3/BiFeO_3$ multilayers (Figure 3e).[158] This modified the metallic $SrRuO_3$ layer. Mao et al. further demonstrated electrical field control of the Ru ion displacement via switchable ferroelectric polarization in a $SrRuO_3/BaTiO_3$ superlattice using in situ transmission electron microscopy, as shown in **Figure 3f**.[112] By applying an electrical field along the out-of-plane direction, they controlled the ion displacement in both the atomically thin metallic $SrRuO_3$ and ferroelectric $BaTiO_3$ layers within the superlattice. They suggested that the asymmetry at the interface due to BaO-$RuO_2$ and SrO-$RuO_2$ can induce polarization bistability along the upward and downward directions, leading to the electrical field control of the displacement of Ru ions. On the other hand, finite octahedral rotation led to the two perovskite unit cell periodicity (unit cell doubling) and the resultant half-order Bragg diffraction peak, as studied by May et al.[166] **Figure 3g** shows the octahedral distortion variation in the $SrRuO_3/SrTiO_3$ superlattice observed by half-order diffraction measurements using synchrotron X-ray facilities.[107]

### *3.1.2 Crystalline Symmetry*

The $RuO_6$ octahedral distortion determines the crystalline symmetry of $SrRuO_3$ based superlattices. Off-axis X-ray diffraction measurements around (204) $SrTiO_3$ Bragg reflections with four different azimuth angles ($\phi$) of 0°, 90°, 180°, and 270° typically revealed the asymmetric crystal structure of orthorhombic $SrRuO_3$ heterostructures (**Figure 4a**).[49] In conventional $SrRuO_3$ single films, the crystalline symmetry of $SrRuO_3$ is modulated by controlling the thickness,[68] epitaxial strain,[69] and chemical composition of the $SrRuO_3$ layers.[64] For superlattices, the control of octahedral tilt propagation was shown to result in an orthorhombic to tetragonal structural phase transition in $SrRuO_3/SrTiO_3$ superlattices by manipulating the thickness of the $SrTiO_3$ layer (**Figure 4b**).[91] A structural phase map as a function of both $SrRuO_3$ and $SrTiO_3$ layer thicknesses was constructed, as shown in **Figure 4c**. Despite the same thickness and stoichiometry of the $SrRuO_3$ layer, the interfacial octahedral connectivity between $SrRuO_3$ and $SrTiO_3$ with atomic-scale thickness modulation systematically realized the structural phase transition of the $SrRuO_3/SrTiO_3$ superlattice. This further modified the electromagnetic properties of the ferromagnetic $SrRuO_3$ layer, including the magnetic coercive field, saturation magnetization, and ferromagnetic transition temperature. Sahoo et al. confirmed similar structural phase transitions in $SrRuO_3/PrMnO_3$ superlattices by varying the thickness of the $PrMnO_3$ layer, as shown in **Figure 4d**.[147] A shift in the magnetic hysteresis loop was observed, and it was enhanced in the case of the tetragonal superlattice, thus indicating crystalline symmetry-dependent interfacial antiferromagnetic coupling. The distorted lattice structure in the $SrRuO_3$ layer within the superlattice can also influence the crystalline symmetry of the partner layer, with inversion symmetry breaking at the interface. Behera et al. proposed a possible orthorhombic structure in the $La_{0.7}Sr_{0.3}MnO_3$ layers within $SrRuO_3/La_{0.7}Sr_{0.3}MnO_3$ superlattices on a (110)-oriented $SrTiO_3$ substrate using X-ray diffraction and Raman spectroscopy (**Figure 4e**).[129] Four unit cell layers of $La_{0.7}Sr_{0.3}MnO_3$, sandwiched between 14 unit cell layers of $SrRuO_3$, exhibited an orthorhombic structure, leading to modification of the magnetic behavior. Using density functional theory calculations and ferroelectric measurements, Callori et al. discussed the possibility of three different types of interfaces depending on the chemical composition and inversion symmetry breaking in the $SrRuO_3/PbTiO_3$ superlattice (**Figure 4f**).[41] As the thickness of the $PbTiO_3$ layer within the superlattices decreased, the polarization asymmetry was enhanced owing to the compositional inversion symmetry breaking, which led to modified ferroelectric phases. Hence, the superlattice approach provides facile and versatile controllability of the bonding geometry in correlated quantum oxides, thus providing numerous intriguing functionalities.

### 3.2 Correlated Phonon Engineering

Facile structural modification in artificial quantum oxides enables the manipulation of phonon dispersion, phonon group velocity, electric polarization, and the density of states of phonons.[167] This has potential for future acoustic applications, including quantum Bragg mirrors and cavities,[168] quantum acoustic memory and transducers,[169] microwave-optical converters,[170] quantum amplifiers,[171] and circuit acoustodynamics.[172] Most studies on phonon engineering have been limited to high-quality III–V compound semiconductors.

However, recent phonon engineering in correlated oxides can lead to novel applications of tunable coherent phonon excitations for future quantum acoustic devices integrated with their correlated functionalities.

*3.2.1 Zone-Folded Phonon and Its Dynamics*

The superimposed periodicity of the superlattice induces the backfolding of the acoustic phonon dispersion, thereby generating zone-folded acoustic phonons in the GHz and/or THz frequency range. Jeong et al. demonstrated phonon dispersion in $SrRuO_3/SrTiO_3$ superlattices, as shown in **Figure 5a**.[88] As the supercell periodicity is designed to be ten times of the bulk $SrTiO_3$ lattice constant, the reciprocal lattice constant of the superlattice decreases by ten times with the appearance of an equivalent backfolded Brillouin zone. This leads to zone-centered acoustic phonons near the Γ-point. The left panel of **Figure 5b** shows the theoretically estimated phonon dispersion with the presence of zone-folded acoustic phonons in $SrRuO_3/SrTiO_3$ superlattices. Clear Raman excitations of zone-folded acoustic phonons with systematic control of their excitation frequency in the THz frequency range were demonstrated via precise thickness control of the supercell periodicity (middle and right panels of Figure 5b). Such coherent phonon modes of superlattices were also observed in the time-resolved pump-probe measurement, as shown the top panel of **Figure 5c**, studied by Yang et al.[103] By examining the fast Fourier transformed pump-probe spectra of the $SrRuO_3$/($SrTiO_3$ or $SrIrO_3$) superlattices, the excitation frequency and mean free path of the zone-folded acoustic phonons were obtained. The bottom panel of Figure 5c shows the experimentally observed phonon mean free path following the theoretical model, thus validating the experimental approaches. Bojahr et al. compared the oscillation phase of zone-folded acoustic phonons in optical pump-probe measurements to ultrafast X-ray diffraction in $SrRuO_3/SrTiO_3$ superlattices (**Figure 5d**).[108] Both optical measurements showed the same fluence dependence of the phase but complex wavelength dependence of the broadband data. Schmising et al. reported the ultrafast structural evolution of the lattice and polarization dynamics in a $SrRuO_3/PbZr_{0.2}Ti_{0.8}O_3$ superlattice using femtosecond X-ray diffraction.[113] **Figure 5e** shows the relative angular shift of the selected Bragg reflection, representing a transient expansion of the superlattice structure by 0.24%. The optical excitation of the $SrRuO_3$ metal layers generates a large stress that compresses the $PbZr_{0:2}Ti_{0.8}O_3$ layers within the superlattice, thereby offering optically controllable ferroelectric polarizations. On the other hand, the photoinduced ultrafast stress generation in ferromagnetic $SrRuO_3$ confirmed the subpicosecond magnetostriction effect in the nanolayered $SrRuO_3/SrTiO_3$ superlattice via ultrafast X-ray diffraction.[95] It was shown that the phonon-mediated and magnetostrictive stress contributions had similar strengths but opposite signs in the $SrRuO_3/SrTiO_3$ superlattice. The amplitudes of the ultrafast X-ray signals (solid diamonds) followed the temperature-dependent magnetization squared below the ferromagnetic transition temperature. This represents a close interplay between the lattice and spin degrees of freedom in $SrRuO_3$ heterostructures.

### *3.2.2 Spin-Phonon Coupling*

Strong spin-phonon coupling plays an important role in understanding the correlated long-range spin order in $SrRuO_3$ systems. For example, in 1994, Kirillov et al. reported temperature-dependent Raman spectra of a $SrRuO_3$ single film to exhibit phonon anomalies at the ferromagnetic transition temperature (**Figure 6a**).[173] The report However, they could not investigate the polarization selection rules for the phonon analysis owing to the multidomain structure of $SrRuO_3$ single films. Iliev et al. further developed the polarized Raman spectra of $SrRuO_3$ films with a thickness of approximately 300 nm grown on (010)- and (001)-oriented $SrTiO_3$ substrates (**Figure 6b**).[174] The authors tentatively assigned the experimentally observed Raman excitations using lattice dynamical calculations. **Figure 6c** summarizes the temperature dependence of the peak position and linewidth of the four different Raman branches of the (010)-oriented $SrRuO_3$ film. The temperature dependence of the phonon peak position exhibited an anomaly around the ferromagnetic transition temperature. The temperature anomaly of the 391 $cm^{-1}$ phonon mode (regarding apical oxygen vibrations) could be understood as spin-exchange coupling modulated by ionic displacement. However, other phonon anomalies (especially for the 225 $cm^{-1}$ phonon mode is mainly related to Sr ion vibration) could not be explained by the conventional magnetostriction effect. This suggests that the mechanism of spin-phonon coupling in $SrRuO_3$ is complicated. Although the phonon linewidths have a relatively large experimental error, the authors suggested that the linewidth peaked near the transition temperature, which also supports the coupling between the optical phonons and fluctuations of the magnetic order parameter near the phase transition.

A more recent study on $SrRuO_3/SrTiO_3$ superlattices has demonstrated that a similarly strong spin-phonon coupling can be preserved in the atomically thin $SrRuO_3$ layers within the superlattice, as shown in **Figure 6d**.[70] A slight difference in the ferromagnetic transition temperature between the single film and superlattice was detected in the phonon spectra, thus providing experimental proof of the preserved spin-phonon coupling. Furthermore, the temperature-dependent Raman spectra of the $SrRuO_3/SrTiO_3$ superlattice shows that the phonon mode at ~367 $cm^{-1}$ split into two modes, at ~358 and ~386 $cm^{-1}$ below the ferromagnetic phase transition temperature (**Figure 6e**). The split in phonon mode could be assigned to oxygen vibrations with orthogonal polarizations in orthorhombic bulk $SrRuO_3$ by utilizing lattice dynamical calculations. It was proposed that the superposition of two orthogonal linear phonon modes with phase differences could give rise to circular phonons, which could combine with spins or magnetic moments.[55] The molecular field in the ferromagnetic phase of the $SrRuO_3$ layer induced the phonon Zeeman effect and the resultant temperature-dependent phonon splitting below the ferromagnetic transition temperature in the Raman spectra, thus evidencing chiral symmetry breaking. This revealed a chiral phonon–mediated interlayer exchange interaction and synthetic spin structure in the superlattice. The synthetic spin structure will be further discussed in Section 3.5.2.

### 3.3 Tunable Ferromagnetism

Superlattice design provides unprecedented tunability of spin ordering in $SrRuO_3$ layers. This Section discusses various magnetic and magnetotransport behaviors in $SrRuO_3$ based superlattices. We also introduce unconventional magnetic interactions across the heterointerface and/or between $SrRuO_3$ layers within the superlattices. This summary can inspire the design of correlated magnetic superlattices for future spintronic applications. These applications may include magnetoresistive memories, magnetic racetracks, spin–orbit torque memories, and current-induced magnetization switching devices.[60, 175, 176]

#### *3.3.1 Structural Modulation of Ferromagnetic Order*

Owing to the finite spin-orbit interaction, the global and local structural modifications of the $SrRuO_3$ layer lead to tunable anisotropic ferromagnetic properties, including saturation magnetization, coercive field, and magnetotransport behaviors. For example, Jeong et al. examined the crystalline symmetry-dependent spin state of a $SrRuO_3/SrTiO_3$ superlattice via atomic-scale thickness control of the $SrTiO_3$ layers.[91] As the orthorhombic distortion of the $SrRuO_3$ layer evolves toward tetragonal symmetry, the crystal field splitting between the $d_{xy}$ and $d_{xz,yz}$ orbitals is enhanced, resulting in a reduction in saturation magnetization. This is schematically shown in the left panel of **Figure 7a**. X-ray abortion spectroscopy with linear dichroism demonstrated the enhancement of the unoccupied state at $d_{xy}$ orbitals of the tetragonal $SrRuO_3/SrTiO_3$ superlattice (right panel of Figure 7a), experimentally supporting this argument. **Figure 7b** shows the structural dependence of magnetization as a function of the magnetic field for the orthorhombic and tetragonal $SrRuO_3/SrTiO_3$ superlattices. The magnetic coercive field significantly increased for the tetragonal superlattices compared with the orthorhombic ones. Lin et al. consistently demonstrated that $SrCuO_2/SrRuO_3/SrCuO_2$ heterostructures exhibits structure-dependent magnetic hysteresis behavior via dimensionality control of infinite-layer-$SrCuO_2$ (**Figure 7c**).[117]

Magnetotransport measurements also reveal the structural modulation of the ferromagnetic behavior. Evidently, the magnetotransport signal originates predominantly from the metallic $SrRuO_3$ layer when the partner layer of the superlattice is insulating. **Figure 7d** shows the magnetic field-dependent magnetoresistance (left panel) and anomalous Hall effect (right panel) of the $SrRuO_3/SrTiO_3$ superlattices, confirming the structural evolution of the magnetic coercive field depending on the $SrTiO_3$ layer thickness.[89] Furthermore, an anomalous Hall signal contains contributions from topological spin dynamics. Chen et al. theoretically suggested the existence of Weyl nodes in ferromagnetic $SrRuO_3$, inspired by the observation of nonmonotonic temperature dependence of anomalous Hall signal and magnetization. These observations suggest a finite Berry curvature in the $SrRuO_3$.[177] Itoh et al. measured the temperature-dependent spin-wave gap using neutron diffraction, exhibiting identical nonmonotonic temperature dependence. This further supports the presence of Weyl Fermions along with spin-wave excitation.[81] Based on these studies, Jeong et al. showed that the size of the spin-wave gap could be further enhanced by using a tetragonal $SrRuO_3$ layer, induced by the modulated magnetic anisotropy in superlattices (**Figure 7e**).[89] **Figure 7f** shows the temperature-dependent anomalous Hall resistivity of the $SrRuO_3/Pr_{0.7}Ca_{0.3}MnO_3$ superlattice.[141] Whereas the anomalous Hall signals of orthorhombic $SrRuO_3$ single films (green and yellow lines) and $SrRuO_3/SrTiO_3$ superlattices (red and blue lines) were negative at low temperatures, the tetragonal $SrRuO_3/Pr_{0.7}Ca_{0.3}MnO_3$ superlattice (navy line) exhibited a positive sign. Ziese et al. performed angle-dependent magnetoresistance measurements, as shown in **Figure 7g**.[138] The experimental results for orthorhombic (left panel of Figure 7g) and tetragonal (right panel of Figure 7g) superlattices show good agreement with the fitting results, thus representing the structural dependence of the magnetotransport behavior. Cui et al. reported that the angle-dependent magnetoresistance of $SrRuO_3/SrTiO_3$ superlattices changes from twofold to fourfold perpendicular symmetry with increasing $SrTiO_3$ layer thickness owing to structural modifications (**Figure 7h**).[107]

*3.3.2 Interfacial and Interlayer Exchange Coupling*

Interfacial exchange coupling of the $SrRuO_3$ layer with different magnetic compounds and/or interlayer exchange coupling between the $SrRuO_3$ layers within superlattices leads to an intriguing correlated magnetic order with functionalities. Qu et al. showed the reorientation of lateral magnetic anisotropy in $SrRuO_3/La_{0.67}Ca_{0.33}MnO_3$ superlattices, as shown in **Figure 8a**.[145] The magnetic field-dependent magnetization curves of the superlattices exhibited a rotation of the magnetic easy axis from the orthorhombic [010] to [100] axis with increasing repetition number of the superlattices. By performing X-ray absorption spectroscopy and structural characterizations, the combination of epitaxial strain and interfacial coupling was suggested to be the origin of magnetic easy axis modulation. On the other hand, the exchange interaction across the interface of ferromagnetic $SrRuO_3$ and antiferromagnetic layers frequently revealed an exchange bias effect, characterized by the shift of the magnetic hysteresis loop. Padhan and Prellier reported a large difference between the field-cooled and zero-field-cooled magnetic hysteresis loops and anisotropic behavior depending on the magnetic field direction for $SrRuO_3/SrMnO_3$ superlattices with different $SrMnO_3$ layer thicknesses (**Figure 8b**).[151] Here, anisotropic interfacial coupling was demonstrated between the biasing moments of the ferromagnetic $SrRuO_3$ layer and pinning moments of the antiferromagnetic $SrMnO_3$ layer within the superlattices. **Figure 8c** shows the temperature-dependent exchange bias in the $SrRuO_3/LaNiO_3$ superlattices investigated by Liu and Ning.[157] The appearance temperature of the magnetic bias effect is defined, which is much lower than the magnetic transition temperature of $SrRuO_3$. Singh and Chen reported a vertical shift of the hysteresis loop in $SrRuO_3/BiFeO_3$ superlattices, as shown in **Figure 8d**.[159] This shift was attributed to pinned $Ru^{4+}$ moments at the interface and local defects in the $SrRuO_3$ layer. Ziese et al. reported antiferromagnetic interlayer coupling in $SrRuO_3/La_{0.7}Sr_{0.3}MnO_3$ superlattices.[136] The left panel of **Figure 8e** shows the out-of-plane (circles) and in-plane (squares) magnetic moments as functions of temperature. By varying the $SrRuO_3$ layer thickness, a change in the magnetization below the ferromagnetic transition temperature of $SrRuO_3$ was evident, thus indicating an interlayer-coupled spin order. The magnetic-field-dependent magnetization curves in the right panel of Figure 8e support the $SrRuO_3$ thickness-dependent interlayer coupling in the $SrRuO_3/La_{0.7}Sr_{0.3}MnO_3$ superlattices. Sahoo et al. also studied interlayer exchange coupling in $SrRuO_3/PrMnO_3$ superlattices.[147] This superlattice is composed of two ferromagnets in which the two layers exchange couples. As an example, **Figure 8f** shows the modulated magnetic field-dependent magnetization curves of the superlattice as a function of the $PrMnO_3$ thickness, indicating the interlayer coupling. More recently, Jeong et al. suggested unconventional interlayer exchange coupling and resultant synthetic spin structures via chiral phonons in $SrRuO_3/SrTiO_3$ superlattices.[55]

### 3.4 Dimensionality-Induced Phase Transition

Dimensional crossover with an electromagnetic phase transition is a key concept for realizing and understanding the low-dimensional correlated functionality of $SrRuO_3$ heterostructures. In this section, we introduce the atomic-scale modulation of electronic ground states, which are strongly coupled to the magnetic ground state in $SrRuO_3$ based superlattices. As the dimensionality of $SrRuO_3$ decreases, the ferromagnetic metallic ground state in the bulk becomes unstable, which is indicative of an intriguing dimensional crossover. Innovation in atomic-scale epitaxy and microscopy techniques has opened a pathway for investigating the monolayer limit of $SrRuO_3$ heterostructures. In particular, one advantage of periodic superlattice structures is that they provide measurable experimental signals of monolayer $SrRuO_3$ with tunable functionality compared with $SrRuO_3$ single films. We also summarize the recent debate on the magnetic and transport behavior of monolayer $SrRuO_3$ within superlattices.

#### *3.4.1 Atomic Scale Control of Electronic Ground State coupled Magnetic Behavior*

Ueda et al. examined the magnetic and electrical transport properties of $SrRuO_3/BaTiO_3$ superlattices grown on $SrTiO_3$ (110) substrates with various stacking periodicities.[111] As the periodicity of the superlattice decreases, the temperature-dependent resistivity exhibits an electrical phase transition from metal to insulator owing to the reduced dimensionality of the $SrRuO_3$ layer (top panel of **Figure 9a**). The magnetization versus temperature curves in the bottom panel of Figure 9a shows a concomitant decrease of magnetization and ferromagnetic transition temperature with decreasing periodicity. The inset of Figure 9a further signifies the antiferromagnetic-like behavior of the superlattice consisting of two unit cell layers of $SrRuO_3$ and five unit cell layers of $BaTiO_3$, similar to layered ruthenates. Liu et al. reported that $SrRuO_3/LaAlO_3$ superlattices with varying $SrRuO_3$ layer thicknesses preserved the ferromagnetic behavior down to two unit cell layers of the $SrRuO_3$ with a ferromagnetic transition temperature of 110 K (**Figure 9b**).[116] Izumi et al. demonstrated $SrRuO_3$ thickness-dependent ferromagnetic behavior observed by temperature-dependent magnetization measurement in **Figure 9c**, in $SrRuO_3/SrTiO_3$ superlattices.[86, 87] As the $SrRuO_3$ layer thickness decreased, the magnetic transition temperature systematically decreased to 60 K for the two unit cell layers of $SrRuO_3$ within the superlattice. The monolayer $SrRuO_3$ did not exhibit magnetic transition down to 2 K. Temperature-dependent resistivity curves of the $SrRuO_3/SrTiO_3$ superlattice consistently exhibited a decrease in temperature with maximum magnetoresistivity related to the ferromagnetic transition temperature, with decreasing $SrRuO_3$ layer thickness (**Figure 9d**). In particular, a monolayer $SrRuO_3$ superlattice with five unit cell layers of $SrTiO_3$ showed insulating temperature dependence in the absence of ferromagnetic transitions. More recently, consistent demonstration of dimensional crossover in both electronic and magnetic phase transitions for atomically thin $SrRuO_3$ layers within $SrRuO_3/SrTiO_3$ superlattices was reported.[92] A clear distinction between the metallic and insulating phases was also observed from temperature-dependent thermopower measurements. Suppression of the electronic charge carriers in the monolayer $SrRuO_3$ superlattice induced a large enhancement in the thermopower at low temperatures. Furthermore, a phase instability at the border of the dimensional crossover, i.e., at the two unit cell layers of $SrRuO_3$, with a temperature-dependent metal-insulator transition along with a metamagnetic transition, was observed. Notably, distinct partner oxide layers of two unit cell $SrRuO_3$ superlattices exhibit exotic magnetic behavior with transition temperature, thus indicating the tunable magnetic functionality of the low-dimensional $SrRuO_3$ superlattices. **Figures 9f-9i** further show the dimensionality-induced electronic phase transition of the $SrRuO_3/SrTiO_3$ superlattices with strongly coupled magnetic phase transitions.[54, 98, 100] Numerous experiments by different research groups consistently demonstrated the presence of electromagnetic phase transition across the dimensional crossover in the $SrRuO_3$ layers.

### *3.4.2 Debate on Monolayer $SrRuO_3$ within Superlattices*

The monolayer limit of $SrRuO_3$ has tremendous potential for achieving a novel low-dimensional ground state in the electronic, magnetic, and topological states. Alves et al. theoretically proposed a spin-polarized two-dimensional electron gas in a $SrRuO_3/SrTiO_3$ superlattice composed of a monolayer of $SrRuO_3$ and five unit cell layers of $SrTiO_3$, as schematically illustrated in **Figure 10a**.[102] The authors used local spin density approximation methods with an effective on-site Coulomb interaction of 4.0 eV for the Ru-*d* orbitals. **Figure 10b** shows the typical double peak feature of the Ru-$d_{xz,yz}$ states (marked by arrows) that are similar to those frequently observed in 1D metallic systems, representing the low-dimensional electronic structure of the monolayer $SrRuO_3$ superlattice. Jeong et al. demonstrated the systematic evolution of the dimensionality-induced metal-insulator transition coupled with the magnetic phase transition using the generalized gradient approximation method (**Figure 10c**).[92] Whereas the trilayer $SrRuO_3$ superlattice preserved the ferromagnetic metallic state similar to that of the bulk, the ground states of the monolayer and bilayer $SrRuO_3$ superlattices were antiferromagnetic insulating states with a finite band gap between the $d_{xy}$ and $d_{yz,zx}$ orbitals of the Ru-$t_{2g}$ states. Thus, they suggested that anisotropic hybridization of the Ru-$t_{2g}$ orbitals induced dimensionality-induced concomitant magnetic and electronic phase transitions in the atomically thin $SrRuO_3$ superlattices. It was further mentioned that the on-site Coulomb interaction value is important for theoretically reproducing the experimentally observed electromagnetic phase of low-dimensional $SrRuO_3$ systems.

The experimental realization and characterization of $SrRuO_3$ monolayers have received considerable attention. Compared with single film, monolayer $SrRuO_3$ superlattices have a structural advantage in terms of accessible experimental signals required to explore the low-dimensional $SrRuO_3$ layers. **Figure 10d** shows the clear insulating behavior of the monolayer $SrRuO_3$ superlattice, characterized by Deng et al.[54] Note that the resistivity value was much lower compared with that for the $SrRuO_3$ single film of three atomic unit cell thickness, in which the thinner films could not be measured.[178] The superlattices were also beneficial for preserving the layer-by-layer growth of the $SrRuO_3$ layer by controlling laser fluence of pulsed laser epitaxy,[54] with minimized surface depletion effect. However, we also note that the quantitative resistivity values of monolayer $SrRuO_3$ superlattices reported by different groups show considerable differences, suggesting the highly sensitive electrical transport behavior of these superlattices within different dielectric environmentsthat mostly provided by the neighboring $SrTiO_3$ layers. Boschker et al. reported monolayer $SrRuO_3$/$SrTiO_3$ superlattices with five unit cell layers of $SrTiO_3$.[105] **Figure 10e** shows the temperature-dependent resistivity of three different superlattices with the same periodicity, revealing diverse resistivity values, particularly at the low temperatures. **Figures 10f** and **10g** consistently show the clear insulating behavior of the monolayer $SrRuO_3$ in other $SrRuO_3$/$SrTiO_3$ superlattices, where the resistivity significantly increases with the thickness of the $SrTiO_3$ layer, as observed by two different groups.[99, 107]

To confirm the effect of the thickness of the $SrTiO_3$ layer in the monolayer $SrRuO_3$ superlattices, we summarized the reported temperature-dependent resistivity of the $SrRuO_3$ monolayer within different $SrRuO_3$/$SrTiO_3$ superlattices (**Figures 10h** and **10i**).[86, 92, 99, 105, 107, 109] Figure 10h shows a large variation in the experimental observations from different groups for monolayer $SrRuO_3$ superlattices depending on the thickness of the $SrTiO_3$ layer. The resistivity values extracted at 40 and 300 K as a function of $SrTiO_3$ layer thickness are shown in Figure 10i. Evidently, most monolayer $SrRuO_3$ superlattices exhibited increased resistivity as the $SrTiO_3$ layer thickness increased at both the selected temperatures. This simple analysis indicates a viable interlayer electronic coupling between the $SrRuO_3$ monolayers through an atomically thin $SrTiO_3$ layer, thus determining the electronic state of the monolayer $SrRuO_3$ superlattices. However, so far, several groups have succeeded in depositing $SrRuO_3$ monolayer superlattices and reported only preliminary experimental results. We hope that future studies will investigate how the interlayer coupling between $SrRuO_3$ monolayers determines the magnetic ground state associated with the metal-insulator transition. Furthermore, most studies so far have been limited to $SrRuO_3$/$SrTiO_3$ superlattices, where charge transfer at the interface is highly suppressed.[92] The suppression of charge transfer at the $SrRuO_3$/$SrTiO_3$ interface of has been predicted theoretically by Zhong and Hansmann, in which the energy level of oxygen ion at the interface has been aligned to predict the difference in work function or electron affinity.[179] $SrRuO_3$ and $SrTiO_3$ happen to have similar energy level of oxygen *p*-orbital state with respect to the Fermi energy. This result suggests that a different partner layers other than the $SrTiO_3$ layer might lead to non-zero charge transfer at the interface, thereby resulting in novel electronic phases in the monolayer $SrRuO_3$ superlattices. **Figure 10j** shows the temperature-dependent resistivity curves of superlattices with monolayer $SrRuO_3$ and 10 unit cell layers of $BaTiO_3$, which is more metallic than the more extensively studied $SrRuO_3$/$SrTiO_3$ superlattices.[112] This experimental observation implies that more intriguing physics might be hidden in the monolayer $SrRuO_3$ superlattices.

### 3.5 Synthetic Spin Structures

Owing to the strongly correlated electrons and aconsiderable spin-orbit interaction, $SrRuO_3$ based superlattices have introduced various novel synthetic spin structures. In this Section, we discuss synthetic spin structures via spin-phonon coupling and topologically nontrivial spin textures observed and manipulated in $SrRuO_3$ based superlattices.

#### *3.5.1 Spin-Phonon Coupling Induced Synthetic Spin Structures*

As mentioned in Section 3.3.2, ferromagnetic $SrRuO_3$ is known to have strong spin-phonon coupling, which has been experimentally validated for both single films and superlattices. Based on the strong spin-phonon coupling, the existence of chiral phonons in the $SrRuO_3$/$SrTiO_3$ superlattice was proposed based on the observation of the phonon Zeeman effect.[55] Chiral symmetry breaking of phonons induces emergent quantum magnetic phenomena such as the phonon Hall effect,[180] optically driven effective magnetic field,[181] AC Stark effect,[182] topologically induced viscosity split,[183] and pseudogap phases.[184] Jeong et al. introduced a chiral phonon-induced interlayer exchange interaction between ferromagnetic $SrRuO_3$ and the resultant synthetic spin structures in $SrRuO_3/SrTiO_3$ superlattices (**Figure 11a**). The left panel of Figure 11a shows the unexpected oscillatory behavior of in-plane magnetization as a function of the $SrTiO_3$ layer thickness in the $SrRuO_3/SrTiO_3$ superlattices. The noncollinear spiral spin state observed from the polarized neutron reflectivity was suggested to be responsible for the observed magnetic oscillation (middle and right panels of Figure 11a), thus suggesting the presence of interlayer exchange coupling. As $SrTiO_3$ is insulating, the conventional Ruderman-Kittel-Kasuya-Yosida interaction cannot explain the observed behavior. Instead, chiral phonons have been proposed as mediating quasi-particles for unconventional interlayer-exchange interactions.

### *3.5.2 Topologically Nontrivial Synthetic Spin Structures*

The strong antiferromagnetic interlayer coupling and/or Dzyaloshinskii-Moriya interactions at the symmetry-broken interfaces of the superlattice can lead to a topologically nontrivial synthetic spin texture, i.e., magnetic skyrmions, in $SrRuO_3$, envisioning novel spintronic functionalities. Noncollinear spin structures in $SrRuO_3$ based superlattices are typically manifested by the topological Hall effect in magnetotransport measurements. Vrejoiu and Ziese proposed that strong antiferromagnetic interlayer coupling with different magnetocrystalline anisotropies in $SrRuO_3/La_{0.7}Sr_{0.3}MnO_3$ superlattices can induce the topological Hall effect, as shown in **Figure 11b**.[122] In these superlattices, the exchange field was estimated to be in the range of several Tesla. Because the ferromagnetic $La_{0.7}Sr_{0.3}MnO_3$ layer is comparatively soft, the topological Hall effect can develop in the $SrRuO_3$ layers. Yao et al. observed a temperature-dependent topological Hall effect in $SrRuO_3/BiFeO_3$ multilayers originating from the broken inversion symmetry near the interface (**Figure 11c**).[158] Jeong et al. investigated the tunable topological Hall effect in a $SrRuO_3/SrTiO_3$ superlattice by controlling the repetition number of the supercell structures, as shown in **Figure 11d**.[90] The competition between the Dzyaloshinskii-Moriya and dipole-dipole interactions can be modulated by the repetition number of $SrRuO_3/SrTiO_3$ superlattices. Using Monte-Carlo simulations, it was shown that the Dzyaloshinskii-Moriya interaction-stabilized Néel-type skyrmions for low-repetition superlattices and dipole-dipole interaction-stabilized Bloch-type skyrmions for high-repetition superlattices were realized. This resulted in a nonmonotonic dependence of the topological Hall signal depending on the repetition number. By adopting a strong spin-orbit coupling of 5*d* oxides with ferromagnetic $SrRuO_3$, Pang et al. further reported the interfacial Dzyaloshinskii-Moriya interaction in $SrRuO_3/SrIrO_3$ superlattices, resulting in a topological Hall effect (arrow in **Figure 11e**).[163]

It is worth discussing other potential mechanisms of the unconventional Hall signal in $SrRuO_3$ based heterostructures. Kan et al. suggested that the inhomogeneous magnetoelectric properties of atomically thin $SrRuO_3$ layers induce a topological Hall-like hump signal as a function of the magnetic field strength.[185] Yang et al. reported the absence of the hump-like signal in the Hall resistance loops of $SrIrO_3/SrRuO_3/SrIrO_3$ trilayers (top panel of **Figure 11f**), whereas the $SrRuO_3/SrIrO_3$ superlattices exhibited a hump structure in the magnetic field-dependent Hall measurements (bottom panel of Figure 11f).[162] This result was interpreted as the unavoidable inhomogeneity in the $SrRuO_3$ layers within the superlattices. Ziese et al. reported that the hump signal of the $SrRuO_3/Pr_{0.7}Ca_{0.3}MnO_3$ superlattice can be explained by a combination of two different Hall contributions, which have different temperature dependence and opposite Hall sign, as shown in **Figure 11g**.[138] The structural modification of $SrRuO_3/Pr_{0.7}Ca_{0.3}MnO_3$ superlattices separates the tetragonal and orthorhombic structures within the $SrRuO_3$ layers. Based on their structural characterization, they suggest that combining two different magnetic domains from tetragonal and orthorhombic structures, which have opposite Hall signals, could potentially lead to a topological Hall-like hump signal. Clear experimental observations are necessary for a better understanding of the unconventional Hall signal. Several techniques for imaging the topological non-trivial spin textures includes magnetic force microscopy, scanning tunneling microscopy, Lorentz transmission electron microscopy, and magnetic X-ray scattering. First, magnetic force microscopy and scanning tunneling microscopy have limitations in observing the spin texture of the inner $SrRuO_3$ layers in some heterostructures because they use a probe tip. Conventional Lorentz transmission electron microscopy still has limitations in high-magnification measurements at temperatures below 150 K and under a magnetic field of about 1 T. We expect that magnetic X-ray scattering is considered the most realistic method for observing the spin texture of the internal $SrRuO_3$ layers while applying an external magnetic field at low temperatures.[186] If changes in the superlattice diffraction peak are observed, distinguishable observations can be obtained due to the significantly larger experimental signal provided by the superlattice peak compared to that of thin single films.

As exemplified by the abovementioned intriguing Hall effect and the potential existence of real-space or reciprocal-space magnetic textures,[74] heterostructures composed of $SrIrO_3$ and $SrRuO_3$ have attracted intense attention in the past few years. As this portion of the work has been summarized in detail in a recent review, it is skipped in the current work. It is worthy to note that iridate by itself is another good example showcasing how the intriguing interplay of electron correlation and spin orbit coupling leads to a rich variety of emergent phenomena.[187-191] In the next Chapter, we will review studies related to iridate-based superlattices other than $SrIrO_3$/$SrRuO_3$ heterostructures.

## 4. 5*d* $SrIrO_3$ Based Superlattices

$SrIrO_3$ is the most notable building block for constructing 5*d* superlattices because of the readily obtained perovskite structure in thin films. 40 nm is believed to be the critical thickness to keep the high quality of perovskite $SrIrO_3$ thin film, above which impurity phases like monoclinic $SrIrO_3$ and Ir-deficiency phases may exist.[192] Nonetheless, it would be helpful to start with the physics related to $Sr_2IrO_4$, to better understand the physics of the quasi-isolated square lattice composed of four $IrO_6$ octahedra, and to better apprehend the motivation for engineering the square lattice in different superlattice structures.[193]

A well-recognized scenario in iridates is that the octahedral crystal field splits the 5*d* orbitals into $e_g$ and $t_{2g}$ orbitals. The latter entangles with spin, giving rise to a half-filled $J_{eff} = 1/2$ doublet on top of a fully occupied $J_{eff} = 3/2$ quartet. Consequently, although the correlation is much reduced in iridates, it is still strong enough to split the narrow $J_{eff} = 1/2$ band into an empty upper Hubbard band (UHB) and a full lower Hubbard band (LHB), thereby leading to a considerable Mott gap in single-layer iridates, such as $Sr_2IrO_4$.[194, 195] The Mott physics not only accounts for the insulating behavior of $Sr_2IrO_4$ but also renders the square-lattice iridate as another promising candidate for pursuing next-generation high-$T_c$ superconductivity.[196, 197] The $J_{eff} = 1/2$ state also results in novel magnetic properties. The most essential consequence of the strong spin-orbit-entangled state is that simple spin is no longer a good quantum number for describing the electron wave function while the magnetic moments are described within the $J_{eff} = 1/2$ manifold.[198] Nevertheless, the quasi-isotropic shape of the $J_{eff} = 1/2$ state has an equal mixture of $d_{xy}$, $d_{yz}$ and $d_{zx}$ orbitals (under cubic crystal symmetry). Thus, the intersite magnetic interaction on a straight bond is necessarily isotropic, i.e., a Heisenberg exchange interaction is assumed, similar to the $S = 1/2$ sites. This local rotational symmetry thus enables a square lattice of $J_{eff} = 1/2$ states to capture the rich physics of the 2D Hubbard model and 2D Heisenberg model, as schematically shown in **Figure 12**. Previous studies on bulk samples have shed light in this direction but only unveiled the tip of the iceberg owing to the limited flexibility in crystal structure. Table 3 summarizes the reported $SrIrO_3$ based superlattices depending on the functionality of the partner perovskite oxide layer, which fosters the controllability of emergent correlated behaviors.[75, 199-218] Next, we review recent exciting findings on superlattices composed of square-lattice iridates by the virtual extensive control of the lattice structure, chemical composition, etc.

### 4.1 Crystal and Magnetic Structures Manipulation

#### *4.1.1 Manipulating electronic properties via atomic-scale precision control of crystal structure*

Figure 13a shows the representative lattice structure of a $(SrIrO_3)1/(SrTiO_3)1$ superlattice, where perovskite $SrIrO_3$ and $SrTiO_3$ monolayers are stacked alternately along the *c*-axis. The most important building block in the superlattice is a square lattice composed of corner-sharing $IrO_6$ octahedra, whereas the $SrTiO_3$ spacer is cubic and electronically and magnetically inert.[75, 199, 219-223] Similar to all other perovskite systems, $IrO_6$ octahedron may rotate around the *c*-axis, which is dubbed as octahedral rotation, or rotate around the diagonal direction in the *ab*-plane leading to an octahedral tilt, as extensively discussed in the previous Chapter for the ruthenate.[224, 225] Octahedral rotation and tilt play drastically different roles in determining the local/global crystal symmetry of $SrIrO_3$. As schematically shown in **Figure 13b**, for a square lattice with a staggered octahedral rotation pattern, the global rotational symmetry is the same as that without rotation.[226] By contrast, the octahedral tilt breaks not only the local inversion symmetry but also the global rotational symmetry (**Figure 13c**).

Matsuno et al. synthesized a series of $(SrIrO_3)n/(SrTiO_3)1$ superlattices with *n* values ranging from 1 to 4.[75] Electrical transport measurements suggest that $(SrIrO_3)3/(SrTiO_3)1$ is at the boundary between the semimetallic phase and the Mott insulating phase. All the superlattices feature a half-filled $J_{eff} = 1/2$ state thanks to the absence of interfacial charge transfer between Ir and Ti. The degradation of the insulating state with increasing *n* indicates that the electron correlation was successively suppressed due to the increasing $SrIrO_3$ spacer thickness. This control of quantum confinement was later confirmed by Kim et al., based on density functional theory calculations. The study found that the vertical charge hopping within the multiple-layer $SrIrO_3$ spacer increased the $J_{eff} = 1/2$ bandwidth, which in turn reduced the effective electron correlation.[200] The $(SrIrO_3)1/(SrTiO_3)1$ superlattice thus represents as a rare artificial layered structure for simulating the 2D single-band Hubbard model.[201] To introduce additional tunability to the layered structure, Hao et al. systematically changed the $SrTiO_3$ thickness while leaving $SrIrO_3$ only a monolayer thick.[199] The $(SrIrO_3)1/(SrTiO_3)2$ and $(SrIrO_3)1/(SrTiO_3)3$ with thicker $SrTiO_3$ spacers are indeed more insulating than $(SrIrO_3)1/(SrTiO_3)1$ with a $SrTiO_3$ monolayer spacer, similar to the case of $SrRuO_3$ (see Section 3.4.2), indicating that the charge-hopping channel across the $SrTiO_3$ spacer is finite and controllable, although the conduction band of $SrTiO_3$ is approximately 2 eV higher than the upper $J_{\text{eff}}$ =1/2 Hubbard band.[75, 202] The charge-hopping channels across the $SrTiO_3$ in the superlattice structure were reaffirmed in a subsequent theoretical work.[202]

*4.1.2 Magnetic Structure Manipulation*

The superlattice series also exhibits an interesting magnetic phase diagram. While all insulating superlattices are antiferromagnetically ordered at the base temperature, the transition temperature $T_N$ can be changed by over two orders of magnitude.[75] As shown in **Figure 14a**, $(SrIrO_3)1/(SrTiO_3)1$ has a maximum $T_N$ of approximately 150 K, which decreases monotonically with increasing $SrIrO_3$ layer thickness. A well-defined antiferromagnetic transition is almost unobservable in the semimetallic $(SrIrO_3)4/(SrTiO_3)1$. On the other hand, $T_N$ decreased with increasing the $SrTiO_3$ layer thickness, as shown in **Figure 14b**. This difference indicates that the $SrIrO_3$ layer thickness alone does not determine $T_N$ in the superlattices. A possible scenario is that the interlayer coupling decreases by separating the neighboring $SrIrO_3$ layers. In fact, the interlayer coupling is speculated to decrease exponentially with increasing distance from the neighboring magnetic layer.[227] This is consistent with the significantly reduced $T_N$ from ~150 to 40 K when the spatial distance between neighboring $SrIrO_3$ layers increased from a $SrTiO_3$ monolayer to a $SrTiO_3$ bilayer.[199] Additionally, the $T_N$ was found to be approximately the same when increasing the $SrTiO_3$ slab from bilayer to trilayer, which is indicative of reaching the regime where the 2D magnetic ordering is predominantly stabilized by magnetic anisotropy, which will be discussed in more detail in Section 4.3.2.

The magnetic anisotropy in square-lattice iridates is highly sensitive to dimensionality. The magnetization and magnetic scattering measurements suggest that $(SrIrO_3)1/(SrTiO_3)1$ has an easy plane anisotropy,[75] similar to the bulk monolayer-systems $Sr_2IrO_4$ and $Ba_2IrO_4$.[228] To resolve the magnetic anisotropy of $(SrIrO_3)2/(SrTiO_3)1$, Meyers et al. measured the azimuthal angle dependence of the magnetic peak using the magnetic resonant scattering technique.[203] A key difference between the easy-plane and easy-axis anisotropy is that the easy-plane magnetic moments can be completely removed from the scattering plane, giving rise to the zero intensity of magnetic peak, while the magnetic peak intensity of an easy-axis system is nonzero at any finite azimuthal angle.[228-232] In parallel to the easy-axis model, the authors found that the magnetic peak is always observable in the studied azimuthal range. This result demonstrates that $(SrIrO_3)2/(SrTiO_3)1$ has a dominant easy-axis anisotropy,[203] that is similar to that of $Sr_3Ir_2O_7$.

By virtue of the dimensionality controlled magnetic anisotropy, Gong et al. prepared an interesting hybrid superlattice structure with an alternating stacking of $(SrIrO_3)1/(SrTiO_3)1$ and $(SrIrO_3)2/(SrTiO_3)1$ superlattices, as schematically showing in **Figures 14d** and **14e**.[204] It is found that the hybrid superlattice orders only at a single temperature close to the Neel temperature of $(SrIrO_3)2/(SrTiO_3)1$, thus suggesting a uniform magnetic structure. However, detailed magnetic scattering measurements revealed that the hybrid superlattice has a dominant canted easy plane anisotropy similar to that of $(SrIrO_3)1/(SrTiO_3)1$. This result suggests that the magnetic behavior of the hybrid superlattice is not a simple addition to the magnetic order of the monolayer and bilayer superlattices. In contrast, integrating the monolayer member and bilayer member of the Ruddlesden-Popper iridates into a single structure leads to two separate ordering temperatures and the addition of orthogonal anisotropy.[233] The difference between the hybrid superlattice and hybrid Ruddlesden-Popper structure could be ascribed to the extra in-plane half-unit-cell slide between the monolayer and bilayer slabs in the Ruddlesden-Popper structure, which suppresses monolayer-bilayer magnetic coupling.[204] Another possible contribution is that the bilayer in the artificial superlattice is much closer to the spin-flop transition than its Ruddlesden-Popper counterpart.[179]

In particular, square-lattice iridates have magnetic anisotropy primarily owing to the high-order perturbation from the virtual hopping between $J_{\mathrm{eff}} = 1/2$ and $J_{\mathrm{eff}} = 3/2$ states in the presence of a nonzero Hund's coupling. The induced pseudo-dipolar interaction produces a predominantly out-of-plane magnon gap, the sign of which varies with tetragonal distortion. This spin model explains the easy-plane anisotropy of $Sr_2IrO_4$ well and predicts an easy-axis antiferromagnetic state with $IrO_6$ octahedra being heavily elongated along the *c*-axis.[198] However, the predicted magnon gap is much smaller than that measured using resonant inelastic X-ray scattering in $Sr_3Ir_2O_7$.[234] Kim et al. reported that compared to single-layer iridates, an additional interlayer pseudo-dipolar interaction must also be considered in the bilayer system, which eventually leads to a far larger anisotropy term that favors *c*-axis anisotropy irrespective of tetragonal distortion.[231, 234] Meyers et al. found in a superlattice system, the octahedral tilt, albeit small in the $(SrIrO_3)2/(SrTiO_3)1$, significantly reduced the interlayer pseudo-dipolar interaction. Thus, although $(SrIrO_3)2/(SrTiO_3)1$ still hosted *c*-axis anisotropy similar to that of the bulk, it was closer to the phase boundary between the easy-plane and easy-axis magnetic phases, as shown in **Figure 14c**.[203] Nevertheless, Gong et al. found that the octahedral tilt of the hybrid superlattice was smaller than that of $(SrIrO_3)2/(SrTiO_3)1$, indicating that the interlayer pseudo-dipolar interaction is not the primary reason for the dominant easy-plane anisotropy in the bilayer slab. A bilayer-Hubbard model analysis revealed that a moderate correlation strength also leads to a giant magnon gap and suggested that the bilayer system in the intermediate regime could be the long-sought excitonic insulator.[235]

## 4.2 Enhancing and Manipulating $J_{eff} = 1/2$ Pseudospin Fluctuations

### *4.2.1 Extreme Spin Fluctuations in the 2D Limit*

In general, spin fluctuation in the 2D limit is much stronger than that in 3D materials, and the critical behavior of 2D spin fluctuation depends sensitively on magnetic anisotropy.[236, 237] For example, the correlation length diverges in a power-law manner when approaching the ordering temperature in the presence of Ising-type magnetic anisotropy. The divergence is faster if the easy-axis anisotropy is switched to the easy-plane anisotropy. The spin fluctuation then reaches an extreme situation of an exponential-type diverging trend when the anisotropy is zero. According to the Mermin-Wagner theorem, the spin fluctuation of 2D isotropic magnets is so strong that any long-range order will be melted at a non-zero temperature owing to $|M| < \frac{C}{\sqrt{T}} \cdot \frac{1}{\left|\ln\sqrt{|h|}\right|}$, where $M$ represents net magnetization or staggered magnetization, $h$ denotes a uniform field or staggered field in ferromagnetic or antiferromagnetic systems, and $C$ is a constant.[238] Extreme spin fluctuation offers an ideal platform for achieving a highly efficient spin response, which is beneficial for designing ultrafast spintronic devices. Additionally, spin fluctuation in the 2D square lattice is recognized to be closely related to the unconventional superconductivity in cuprates[239, 240] and is essential for advancing fundamental physical understanding.

*4.2.2 Staggered Field Effect in Antiferromagnets*

In certain antiferromagnetic materials with locally broken inversion symmetry, an external magnetic field may behave effectively as a staggered field to linearly couple to the antiferromagnetic order parameter. For example, in Cu benzoate, the Dzyaloshinskii–Moriya interaction can be written in the form $H_{DM} = \sum_j (-1)^j \vec{D} \cdot (\vec{S}_j \times \vec{S}_{j+1})$, where $\vec{D}$ is the Dzyaloshinskii–Moriya vector and $\vec{S}_j$ is the spin operator at site *j*. Here, factor $(-1)^j$ denotes the staggered Dzyaloshinskii–Moriya pattern along the spin chain. After rotation along the $\vec{D}$ direction through an alternating angle α/2, where $\tan\alpha = D/J$, i.e., $\vec{S}_j^+ \rightarrow \tilde{\vec{S}}_j^+ \cdot e^{i\alpha/2}, \vec{S}_{j+1}^+ \rightarrow \tilde{\vec{S}}_{j+1}^+ \cdot e^{-i\alpha/2}$, the Dzyaloshinskii–Moriya term can be absorbed into the exchange interaction, giving rise to an effective Heisenberg-like model.[226] In 2D square-lattice systems, the Hamiltonian transformation can be thought of as the conversion of the global spin frame (**Figure 15a**) to the twisted local frame (**Figure 15b**). Considering the Zeeman effect under an external field $h$ along the *x*-direction $H_{Zeeman} = -h\sum_j S_j^x$ in the redefined spin frame, one will arrive at $H_{Zeeman} = -h\cos\frac{\alpha}{2}\sum_j \tilde{S}_j^x - h\sin\frac{\alpha}{2}\sum_j (-1)^j \tilde{S}_j^y$. Clearly, a staggered field, linearly coupled with the AFM order parameter, is generated along the *y*-axis. Furthermore, the staggered field effect is governed by $D/J$, which generally increases with spin orbit coupling in general. Note that for a 1D spin chain, the staggered pattern of the Dzyaloshinskii–Moriya interaction on neighboring bonds does not break the global rotation symmetry.

*4.2.3 Manipulating 2D Extreme Antiferromagnetic Fluctuations*

The staggered field effect enables direct control of antiferromagnetic fluctuations. As mentioned in Section 4.1.2, $(SrIrO_3)1/(SrTiO_3)2$ and $(SrIrO_3)1/(SrTiO_3)3$ are at the 2D limit because the almost unchanged $T_N$ with further suppressed interlayer coupling.[199, 241] The in-plane staggered rotation pattern of $IrO_6$ octahedra and the large spin orbit coupling gives rise to a substantial staggered field upon application of an in-plane magnetic field. As shown in **Figure 15c**, the crossover temperature $T_0$, which can be treated as the ordering temperature the purpose of practical application, increases rapidly with the in-plane magnetic field. This observation is striking and even counterintuitive, considering that antiferromagnets are generally resilient to external magnetic fields.[242] The giant magnetic response was ascribed to the unique combination of a hidden SU(2) symmetry protected by the global crystal rotational symmetry, a large staggered field effect due to the strong spin orbit coupling, and a vanishingly small interlayer coupling due to the quasi-2D nature of the artificial structure.

The sharp increase in the ordering temperature with the magnetic field indicates that the disordered state above $T_0$ can be transitioned into a long-range order state by applying certain magnetic fields, thus offering a novel approach for designing magnetic field switching of antiferromagnetic order, as shown in **Figure 15d**. The magnetic logic bit (on/off) is typically defined by the orientation of the order parameter in the magnetic ordering state in antiferromagnetic spintronics.[243-246] This is distinct from the above study, where the *on* state does not rely on the specific spin orientation in the ordered state, whereas the *off* state is simply a trivial disordered state. Moreover, the key to realizing such an efficient staggered response to a uniform stimulus is the SU(2)-symmetry-preserved Dzyaloshinskii–Moriya interaction, which can be achieved in various antiferroic materials via sophisticated lattice engineering strategies. This finding may stimulate more exciting work in exploring and realizing advanced nanoscale devices with antiferroic materials, which are much abundant but are largely unexplored.

### 4.3 Enhancing and Manipulating $J_{\rm eff} = 1/2$ Charge Fluctuations

#### *4.3.1 Charge Fluctuations in the Intermediate Regime*

The spin and charge degrees of freedom are entangled in the half-filled single-band Hubbard system because charge localization is tied to the formation of local magnetic moments.[201, 247] Thus, the local magnetic moments are thus effectively local particle–hole pairs. Correspondingly, breaking up a pair by exciting the charges into the electron–hole continuum above the charge gap annihilates the local magnetic moment (i.e., longitudinal spin fluctuations), and vice versa. In the strong-coupling Mott limit, the local magnetic moments survive well above $T_N$ because the charge degree of freedom remains frozen until the thermal fluctuations substantially disturb the Mott gap, which can be several orders of magnitudes higher than $T_N$. Consequently, the spin-charge fluctuation is typically weak in this regime. In the weak-coupling Slater limit, the system is metallic above $T_N$ without the local magnetic moments, thus resulting in weak spin-charge fluctuations. Compared to the two limits, the intermediate coupling regime has an optimum spin-charge fluctuation because both local moments and charge carriers exist above $T_N$. A generic phase diagram of a half-filled single-band system is schematically drawn in **Figure 16a**. Clearly, electron correlation is the key to exploring the emergent phenomena associated with charge fluctuations in a single-band Hubbard system.

*4.3.2 Modulating Electron Correlation via Epitaxial Strain*

Epitaxial strain is one of the most vital and readily accessible knots for modulating the effective electron correlation in thin-film and heterostructure studies. Kim et al. predicted that the $J_{\mathrm{eff}} = 1/2$ bandwidth of the $(SrIrO_3)1/(SrTiO_3)1$ superlattice decreases with increasing tensile strain.[219] Kim *et al.* confirmed the $J_{\mathrm{eff}} = 1/2$ band feature in the superlattice, as shown in Figure 16b, and unveiled that the substantial strain effect in $(SrIrO_3)1/(SrTiO_3)1$ as compared to $Sr_2IrO_4$ arises from the additional vertical hopping channels in the former,[202] as schematically shown in **Figure 16c**. Kim *et al.* later expanded the strain study on $(SrIrO_3)m/(SrTiO_3)1$ with $m > 1$.[200] Ranging from -4% to 4% of the epitaxial strain, the authors found that the electronic structure, albeit slightly different owing to the different stacking patterns of the superlattice series, displayed a similar strain dependence.

Experimentally, Yang *et al.* prepared a series of $(SrIrO_3)1/(SrTiO_3)1$ superlattices on different substrates to implement variable epitaxial strains of 0, -0.95%, and -1.06%.[248] With X-ray absorption spectroscopy measurements at the O *K*-edge, the authors found that the $J_{\mathrm{eff}} = 1/2$ single-band feature was preserved under epitaxial strain. In the above X-ray absorption measurements, the intensity difference between the out-of-plane and in-plane channels was considered to be a measure of the relative hybridization strength of Ir with planar oxygen.[249, 250] Notably, the X-ray absorption intensity in the out-of-plane channel was always larger than the in-plane signal, indicating a dominant hybridization between Ir and planar oxygen in all the superlattices (**Figure 16d**). By extracting the X-ray linear dichroism, which is the difference spectra between the out-of-plane and in-plane channels, the authors unveiled that in-plane hybridization systematically increased with compressive strain. This is strong evidence that the effective electron correlation in superlattices decreases monotonically with increasing compressive strain.

Electronic correlation modulation was also captured in the electric transport and magnetic measurements. Although all the superlattices had an insulating ground state, the insulating behavior was weakened in the compressed superlattices, confirming that the reduced electron correlation is consistent with the electronic structure. Interestingly, the authors observed an emergent high-temperature metallic phase in $(SrIrO_3)1/(SrTiO_3)1$ with the largest compressive strain. Moreover, all the superlattices had the same antiferromagnetic structure while the Neel temperature $T_N$ decreased monotonically with increasing compressive strain. Intuitive, half-filled single-band systems can be ascribed to Mott insulators because of the insulating state being established well above $T_N$.[247] However, the reduced $T_N$, due to enhanced electron hopping, clearly contradicts the expectation of the exchange interaction $J \propto {4t^2}/{U}$ ($t$ and $U$ denote electron hopping and correlation, respectively) at the strong coupling limit.[251] The magnetic peak intensity was quantified at the base temperature and revealed that the antiferromagnetic order parameter decreased monotonically with compressive strain. The fact that the antiferromagnetic order parameter, antiferromagnetic ordering temperature, and strength of the insulating behavior decrease simultaneously with the effective electron correlation (**Figure 16e**), revealed that charge fluctuations play a dominant role in the stability of the antiferromagnetic ordering. This is a key feature of Slater-type insulators in the weak-coupling limit.[247, 252] The coexistence of characteristics of weak and strong coupling limits indicates that the $(SrIrO_3)1/(SrTiO_3)1$ superlattice system should be within the Slater-Mott crossover regime. In contrast to these two limits, knowledge about the intermediate regime is very limited,[252] which is discussed in more detail in the next section.

*4.3.3 Manipulating Charge Fluctuations via Staggered Field Effect*

To explore the intermediate regime in depth, Hao et al. performed a comprehensive magnetotransport study of $(SrIrO_3)1/(SrTiO_3)1$ grown on a $SrTiO_3$ substrate.[253] As mentioned in Section 4.3.1, the intermediate regime (or the Slater-Mott crossover regime) hosts energetic charge fluctuations, which effectively lead to longitudinal spin fluctuations.[248] The intriguing spin-charge interplay causes a notable anomaly in the insulating resistivity when cooling through $T_N$. After applying an in-plane magnetic field, the authors observed a substantial magnetoresistance effect above $T_N$. Positive magnetoresistance is unusual because transverse spin fluctuations typically plays a dominant role in antiferromagnetic semiconductors, and a negative magnetoresistance effect is commonly observed.[254, 255] In addition, magnetoresistance above $T_N$ exhibits a similar temperature dependence of the staggered susceptibility, which increases quickly with temperature approaching $T_N$, as shown in **Figure 16f**. Another interesting observation is the anisotropic characteristic of the anomalous magnetoresistance, i.e., a large magnetoresistance is observed when the magnetic field is parallel to the plane, whereas the out-of-plane magnetoresistance is almost unobservable. Note that because of the finite antiferromagnetic order parameter, a staggered field effect also exists in the intermediate regime, which is expected to stabilize the localized magnetic moments and enhance the particle-hole binding. As a large staggered field is only produced in the presence of an in-plane magnetic field owing to the in-plane net magnetic moment, this observation demonstrates that magnetoresistance originates from the suppression of spin-charge fluctuations under staggered magnetic fields.

Further insights into the anomalous magnetoresistance were gained through theoretical simulation using a 2D model built on a $J_{eff} = 1/2$ square lattice. The authors reported that a staggered field increases the size of the antiferromagnetic order parameter by suppressing longitudinal spin fluctuation. The effect is most significant around $T_N$, where the spin susceptibility is maximized. The carrier density is strongly suppressed owing to the staggered magnetic field and reaches its minimum around $T_N$. The carrier density reduction renders the anomalous magnetoresistance even more peculiar than the carrier mobility-driven magnetoresistance effects in most cases.[256-261] The anomalous magnetoresistance thus probes the longitudinal spin fluctuations by virtue of the spin-orbit coupling-induced staggered field effect, and offers a novel pathway to control the binding energy of the fluctuating particle-hole pairs via a magnetic field in the Slater-Mott crossover regime.

### 4.4 Topology-Correlation Interplay in Artificial Iridates

The intermediate regime is interesting not only because of the coexistence of spin and charge fluctuations but also because of the possibility of integrating both electronic topology and correlation in a single system. The concept of electronic topology was developed with the prediction and discovery of a variety of symmetry-protected surface states, edge states, etc.[262, 263] These studies focused primarily on noninteracting systems that can be well understood in single-particle band pictures. Nonetheless, the interaction between electrons is typically nonzero in real materials and can sometimes lead to emergent phenomenon beyond band structure prediction. A well-known example is the Mott insulating state in 2D half-filled single-band systems, for which non-interacting band theory predicts a good metal. Thus, whether and how the electronic topology manifests in correlated systems is generally unknown. This is particularly true for the intermediate regime, where both the single-electron approach in the weak-coupling limit and the superexchange approach in the strong-coupling limit fail to capture the energetic spin and charge excitations. It is believed that designer electronic topology and electronic correlation based on artificial heterostructures are promising routes to understand the intriguing interplay.[264]

As discussed in Section 4.3.2, $(SrIrO_3)1/(SrTiO_3)1$ is a realistic 2D half-filled Hubbard model system in the intermediate regime. To implement a nontrivial electronic topology, Yang et al. prepared $(SrIrO_3)1/(CaTiO_3)1$ superlattices on $SrTiO_3$ substrates.[206] The key difference between the lattice structures of $(SrIrO_3)1/(CaTiO_3)1$ and $(SrIrO_3)1/(SrTiO_3)1$ is the realization of a large octahedral tilting in the former by reducing the tolerance factor. The effective Hamiltonian of the distorted 2D square lattice can be written as:

$$H = -t \sum_{<ij>} \sum_{\alpha,\beta} \left[ c_{i\alpha}^{\dagger} \left( e^{i\theta \boldsymbol{d}_{ij} \cdot \sigma} \right)_{\alpha\beta} c_{j,\beta} + h.c. \right] + U \sum_{i} n_{i\uparrow} n_{i\downarrow} \quad (1),$$

where the first term accounts for the nearest-neighboring $<ij>$ hopping, as schematically shown in **Figure 17a**, $c_{i\alpha}^{\dagger}$ $(c_{i\alpha})$ is the creation (annihilation) operator of a $J_{\mathrm{eff}} = 1/2$ electron with spin $\alpha$ on site $i$, $\boldsymbol{\sigma}$ is the vector of Pauli matrices, and $U$ is the onsite electron-electron repulsion. In contrast to the typical 2D Hubbard model, electron hopping between neighboring sites is mediated by a finite SU(2) gauge field, $e^{i\theta \boldsymbol{d}_{ij} \cdot \sigma}$. Octahedral rotation contributes to the staggered normal component of the gauge field, whereas octahedral tilt results in planar components. Unlike the rotation, the octahedral tilt breaks the global fourfold rotational symmetry such that the planar components cannot be gauged simultaneously on the square lattice. This effect can be intuitively captured as the gauge-invariant flux $\phi$ of any closed path of hopping, as shown in **Figure 17b**. In the absence of correlation, the model leads to a Dirac semimetal, and the nonzero flux emerges in the momentum space as a Berry phase, which is the main source of the spontaneous Hall effect.[265, 266] In the strong-coupling limit, the flux gives rise to Ising anisotropy along $\boldsymbol{d}_{\square}$.

Consistent with the model analysis, a large anomalous Hall effect below $T_N$ was observed. Interestingly, both the saturated and remnant anomalous Hall conductivities exhibited a nonlinear temperature dependence, i.e., they first increased quickly and then began decreasing with cooling, as shown in **Figure 17c**. This unconventional behavior indicates that the anomalous Hall effect is governed by physical quantities that compete with each other. Indeed, the authors observed that the anomalous Hall angle $\sigma_{yx}/\sigma_{xx}$, which measures the strength of the charge deflection, increased with cooling, whereas the anomalous Hall coefficient $S_H = \sigma_{yx}/M_s$ which is proportional to the effective charge carrier density, had an opposite dependence, as shown in the inset of Figure 17c. Thus, the intrinsic anomalous Hall effect suggests that the impact of electronic topology extends to a Mott insulator in the moderately correlated regime. This is the central idea of the intermediate regime, where delocalized antiferromagnetic moments give rise to activated charge carriers and the two physical quantities increase at the expense of each other.

At the base temperature, the authors found that the magnetic structure indeed had an easy axis along $\boldsymbol{d}_{\square}$, as shown in the inset of **Figure 17d**. Nevertheless, through resonant inelastic scattering measurements, the spin gap was found to be surprisingly large (~85 meV), close to the giant magnon gap of the bilayer iridate $Sr_3Ir_2O_7$.[234] The gap size was far larger than expected from the linear spin wave theory in the strong-coupling limit and can be understood only using a Hubbard model with a moderate correlation strength. The nonmonotonic anomalous Hall effect and the giant spin gap demonstrate that the interplay between the electronic topology and electronic correlation is so rich that the emergent phenomena are typically beyond the theorems relating to the weak-coupling or strong-coupling limit.

In complementary to the intrinsic anomalous Hall effect driven by the same group of $J_{eff} = 1/2$ electrons that form the antiferromagnetic Mott insulating state through symmetry engineering in a 2D square-lattice iridate block, the anomalous Hall effect was also observed in superlattices composed of $SrIrO_3$ and other magnetic compounds, as is discussed in the next section.

### 4.5 Superlattices Composed of $SrIrO_3$ and Magnetic Compounds

The anomalous Hall effect is typically observed in materials with SOC and broken time reversal symmetry.[265] Although $SrIrO_3$ features a large spin-orbit coupling and theorypredicts the existence of multiple nontrivial electronic topologies upon breaking a certain crystal symmetry,[267] the material by itself does not break the time reversal symmetry. Thus, a large effort has been devoted to introducing long-range magnetic order in iridates by heterostructuring $SrIrO_3$ with different magnetic compounds.

For example, Nichols et al. synthesized a series of superlattices composed of $SrIrO_3$ and an antiferromagnetic insulator $SrMnO_3$.[212] In contrast to the two mother phases, the authors found that $(SrIrO_3)1/(SrMnO_3)1$ displayed a large net magnetization with the easy axis along the film normal direction. The interfacial ferromagnetism was ascribed to the charge transfer between Ir and Mn. In parallel to the spontaneously broken time-reversal symmetry, the anomalous Hall effect was observed below the magnetic ordering temperature. As shown in **Figure 18a**, the anomalous Hall effect is reduced and eventually vanished with increasing layer thickness, highlighting the crucial role of the interface in the emergent ferromagnetism. Skoropata et al. replaced $SrMnO_3$ with $LaMnO_3$ in the aforementioned superlattice structures and realized a potential topological Hall effect by engineering the interfacial Dzyaloshinskii–Moriya interaction.[216] Recently, Gu et al. exploited $CaMnO_3$ as the main constituent of a superlattice composed of $SrIrO_3$.[218] The authors observed a metal to nonmetal crossover with decreasing layer thickness (**Figure 18b**), again highlighting the crucial role of the Ir/Mn interface in modulating the global electric transport property. In contrast to building a $SrIrO_3$ based superlattice with magnetic conductors, Jaiswal et al. prepared superlattices with $SrIrO_3$ and $LaCoO_3$, which is a magnetic insulator at low temperatures.[213] They reported a large anomalous Hall effect when $SrIrO_3$ was in directly contacts with $LaCoO_3$ at the atomic scale, whereas such effect was absent when $SrIrO_3$ and $LaCoO_3$ layers were separated by a thick $SrTiO_3$ spacer (**Figure 18c**). The authors argued that the anomalous Hall effect is due to proximity-induced magnetism in the $SrIrO_3$ layer.

In addition to the end members of $R$MnO3, Yoo et al. constructed heterostructures with $SrIrO_3$ and $La_{0.7}Sr_{0.3}MnO_3$,[207] which were best optimized for the double-exchange-dominated ferromagnetism. In addition to the ordinary Hall effect, the authors extracted the anomalous Hall component of the superlattices. With a systematic variation in the layer thickness, they found that the anomalous Hall resistivity always scales linearly with the square of the longitudinal resistivity, as shown in **Figure 18d**.[265] The anomalous Hall effect was ascribed to proximity-induced magnetism in the $SrIrO_3$ layer, whereas the intriguing double-SrO layer at the Ir/Mn interface plays a subtle role in mediating the exchange interaction. Yi et al. found that attaching $La_{0.7}Sr_{0.3}MnO_3$ to $SrIrO_3$ at the atomic scale leads to a large variation in magnetic anisotropy.[208, 210] In addition to $SrIrO_3/La_{0.7}Sr_{0.3}MnO_3$ superlattices, Yi et al. prepared superlattices composed of $SrIrO_3$ and $La_{0.8}Sr_{0.2}MnO_3$.[211] Interestingly, the authors observed a reversible phase transformation with a lattice change as large as 7% at room temperature using ionic liquid gating. The phase transformation is associated with a clear modulation of the electronic, magnetic, and optical properties due to the reversible transfer of oxygen and/or hydrogen ions. This work further demonstrates that digital superlattices usually hosts emergent functionalities that have not been seen in the mother phases or in their solid solutions.

### 4.6 Superlattices Composed of Iridates Other Than $SrIrO_3$

As the 3D end member of the iridate Ruddlesden-Popper family, $SrIrO_3$ has a simple perovskite structure similar to most functional oxides and is widely used to construct iridate-based superlattices. With the development of epitaxy techniques, studies have also been conducted on superlattices composed of other iridates, such as $Sr_2IrO_4$ and $CaIrO_3$.

Liu et al. grew superlattices with alternating $Sr_2IrO_4$ and $SrTiO_3$ layers.[268] The superlattices were found to have an emergent magnetoelectric phase transition, probably due to the double SrO layers at the interface (**Figure 19a**). By replacing $SrTiO_3$ with $BaTiO_3$, the magnetoelectric transition was substantially enhanced, with a pronounced magnetoelectric coefficient. Gruenewald et al. integrated $Sr_2IrO_4$ and $LaSrGaO_4$ atomically in a planar manner and realized 1D $IrO_2$ stripes.[269] From resonant inelastic X-ray scattering measurements, as shown in **Figure 19b**, the authors found that the 1D stripes clearly enhanced exchange interactions and confined orbital excitations compared with $Sr_2IrO_4$. In addition to the interesting dielectric and magnetic behaviors observed on $Sr_2IrO_4$ based superlattices, Xu et al. found an enhanced anisotropic magnetoresistance effect in superlattices composed of $Sr_2IrO_4$ and $La_{0.7}Sr_{0.3}MnO_3$.[270] By modulating the magnetic field direction, a nonvolatile memory based on the AMR effect was realized.

Similar to $SrIrO_3$, $CaIrO_3$ is semimetallic and can be stabilized into the perovskite phase using epitaxy.[271] Recently, $CaIrO_3$ has been increasingly used to construct atomic-scale superlattices. For example, Lim et al. prepared a series of $CaIrO_3$/$SrTiO_3$ superlattices and found enhanced anisotropic magnetoresistance as compared to $SrIrO_3$/$SrTiO_3$ superlattices.[272] Interestingly, the authors found that the anisotropic magnetoresistance of $(CaIrO_3)1/(SrTiO_3)1$ was maximized at 45°, whereas the largest anisotropic magnetoresistance of $(SrIrO_3)1/(SrTiO_3)1$ occurred at 90°. This anisotropic magnetoresistance difference indicates that replacing $SrIrO_3$ with $CaIrO_3$ in the superlattice causes a substantial change in the magnetic structure, as shown in **Figure 19c**. Fourfold anisotropic magnetoresistance was also observed in a recent study by Vagadia et al., who constructed multiple superlattices composed of $CaIrO_3$ and $CaMnO_3$.[273] The anisotropic magnetoresistance magnitude was found to be as high as 70%, which is approximately two orders of magnitude greater than the value reported for $(CaIrO_3)1/(SrTiO_3)1$, as shown in **Figure 19d**. This unprecedentedly high anisotropic magnetoresistance was ascribed to the combination of a strong biaxial anisotropy and a metamagnetic transition.

## 5. Conclusions and Outlook

Ruthenate- and iridate-based oxide superlattices exemplify the power of bottom-up synthesis by heteroepitaxial growth in overcoming the material bottleneck in solid-state chemistry. Unlike the large number of material families of 3*d* transition metal oxides, studies on 4*d* and 5*d* systems have been limited to a few compounds available in bulk form. Artificial crystals built from atomically thin $SrRuO_3$ and $SrIrO_3$ demonstrate a route for creating coherent layered structures that are unobtainable using conventional protocols. A particularly important aspect of the reported studies is that one can create analogous systems to the known bulk compounds. The analogy is not only in the layered character of the crystal structure but also in the physics behind the electronic and magnetic properties. It is crucial to make such comparisons when designing artificial crystals because they determine the crystal unit that play an essential role in the unified picture of the physical properties. A closely related direction is the interface and heterostructures of 3*d* and 4*d*, 3*d* and 5*d*, and 4*d* and 5*d* oxides. One advantage of such systems is that they bring together different degrees of correlation and spin-orbit coupling. Although some studies have adopted this direction, the complexity of such systems often renders difficulty in disentangling the roles of the different layers in driving the observed properties, particularly when the layers are unit-cell thin. The superlattices with quantum-confined structures of atomically thin $SrRuO_3$ and $SrIrO_3$ layers fill the gap in understanding. They established that the ground state and properties of a monolayer of $SrRuO_3$ and $SrIrO_3$ are favorable, which serves as the basis for using them as building blocks for more complex structures. This allows the reexamination of ultrathin superlattices with 3*d* oxides and rejuvenates this research direction. It also sets a blueprint for developing new superlattices of this kind. Note that the analogy between the superlattice and bulk does not mean that the properties of the artificial crystals are the same as those of the bulk in every aspect. Instead, the key is to be able to identify both their similarities and differences and understand the reasons behind them. This builds the foundation for further exploiting the additional structural degrees of freedom afforded by the superlattice to generate structural variants beyond the bulk compounds and stabilize new phases.

An immediate benefit of superlattice synthesis is the possibility of obtaining a large variety of oxide systems with strong spin-orbit coupling, which is the key property that distinguishes 4*d* and 5*d* systems from their 3*d* counterparts. This opens the door to exploring topology-driven physics and properties of oxides through structural and symmetry engineering. This is an important future direction for this class of systems. Proposals of topological designs of oxide superlattices have indeed been put forward for about ten years. However, numerous predictions are based on the single-particle picture, which may not be realistic because of the significant electronic correlation remaining in the 4*d* and 5*d* orbitals. This characteristic could result in 4*d* and 5*d* oxides in a regime where topology and correlation coexist and interact with each other. For instance, one of the most important consequences of correlation is magnetism, and the role of spin-orbit coupling in correlated systems has traditionally been viewed as a perturbation that controls magnetic anisotropy. On the other hand, if spin-orbit coupling leads to a nontrivial electronic topology, magnetically driven topological states may emerge, and their emergence can very well depend on magnetic anisotropy. The intrinsic anomalous Hall effect is an excellent example, in which both the Berry phase and magnetic order with a proper easy axis are necessary. Therefore, one may consider the anomalous Hall effect as a combined result of spin-orbit coupling, correlation, and crystal symmetry, which can be implemented in artificial crystals by design. Even in the strong correlation limit, the impact of topology may still survive but in a different form because the spin-dependent hopping that is responsible for topology is mapped to the Dzyaloshinskii–Moriya interactions. In fact, exchange magnetic anisotropy is a result of Dzyaloshinskii–Moriya interactions, which are a vital source of magnetoelectricity in oxides. Artificial crystals of 4*d* and 5*d* electrons may bridge several different research fields.

From the application point of view, creating magnetic oxides with substantial spin-orbit coupling is crucial for the development of spintronics. Future directions include realizing magnetic materials with especially large magnetic anisotropy for nonvolatile memory and developing efficient spin charge conversion for energy harvesting. Finally, there is great potential for applying new experimental methods from parallel developments. Examples include ionic-liquid gating and freestanding membranes. The superlattice structure could have the advantages of sustaining the lattice when the ions move in and out. Broadly stated, this may have significant implications in catalysis studies because precious metals are known to be good catalysts. Controlling the 4*d* and 5*d* electronic states in artificial crystals can lead to improved performances.

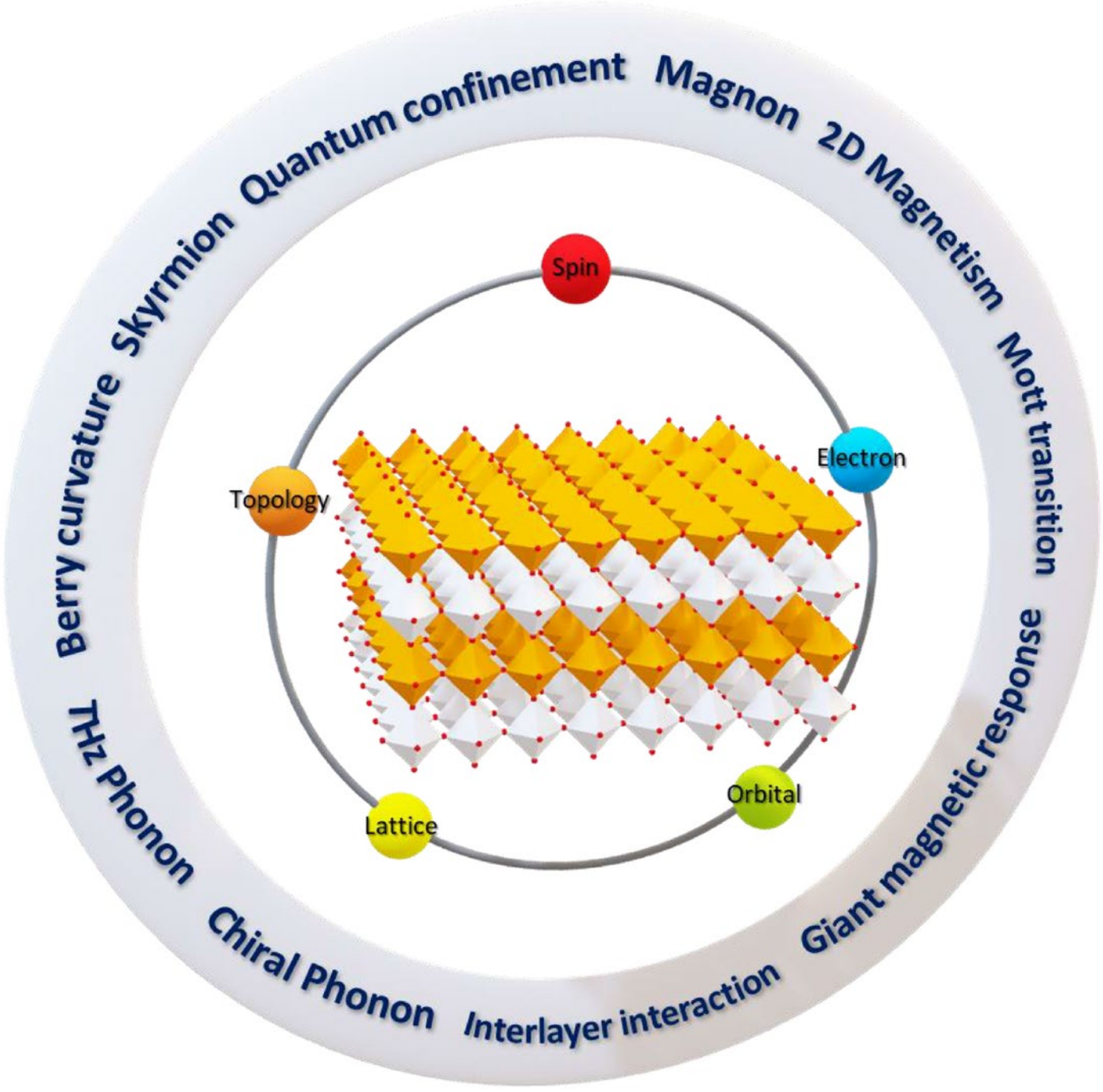

**Figure 1.** Schematic summary of artificial 4*d* and 5*d* oxide superlattices with their versatile functionalities.

**Table 1.** Correlation strength in 3*d*, 4*d*, and 5*d* transition metal oxide superlattices. The symbols $U$, $\lambda$, and $\Delta$ represent on-site Coulomb interaction, spin-orbit coupling, and crystal field splitting energy, respectively.

| Elements | $U$ (eV) | $\lambda$ (eV) | $\Delta$ (eV) |
|---|---|---|---|
| 3*d* | 5 ~ 7 | 0.01 ~ 0.10 | 1 ~ 2 |
| 4*d* | 0.5 ~ 3 | 0.1 ~ 0.3 | 2 ~ 3 |
| 5*d* | 0.4 ~ 2 | 0.1 ~ 1 | 3 ~ 4 |

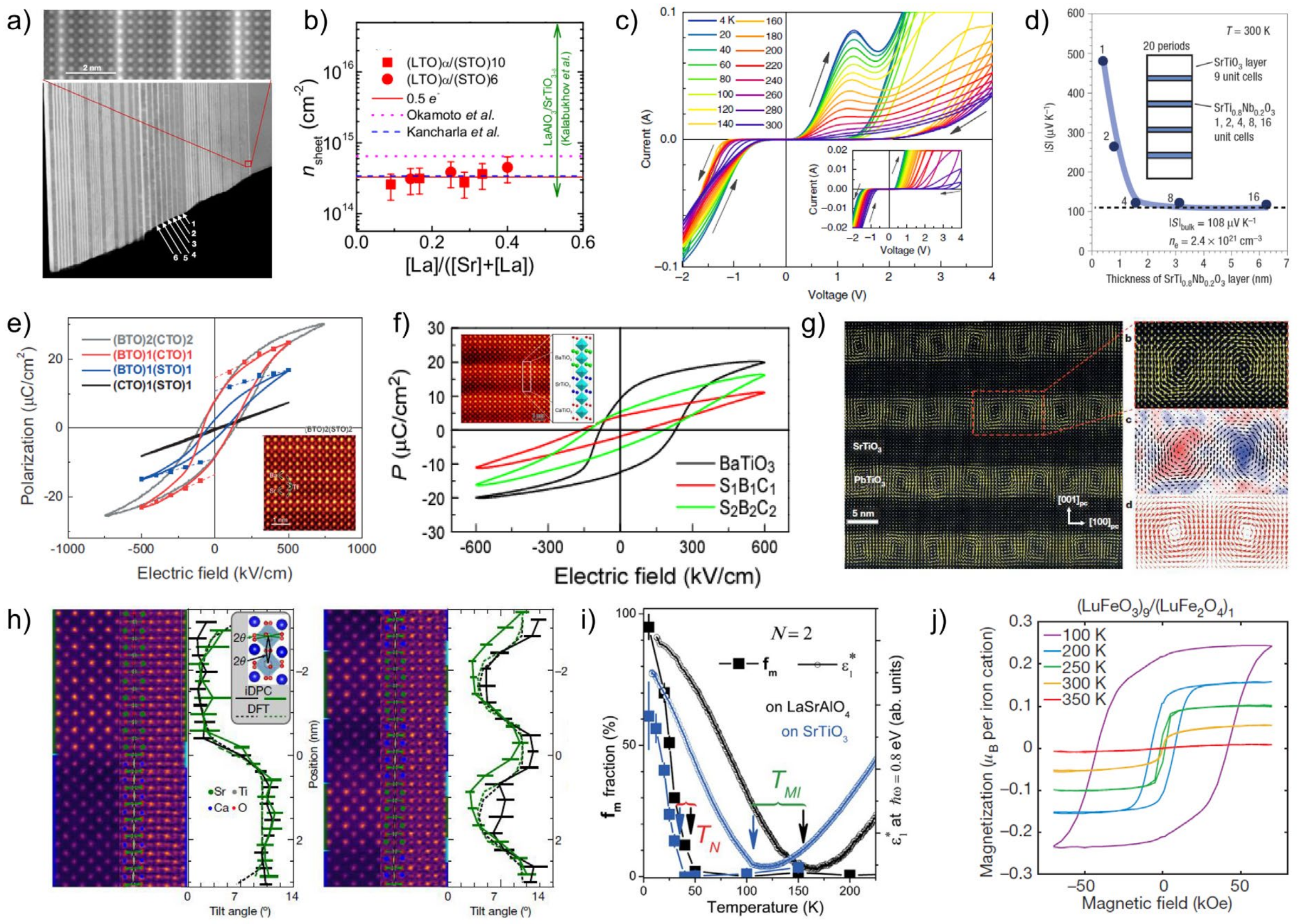

**Figure 2.** a) Cross-section scanning transmission electron microscopy image, reproduced with permission.[16] Copyright 2002, Springer Nature, b) optical spectroscopy analysis, reproduced with permission.[17] Copyright 2007, American Physical Society, and c) quantum resonant tunneling results of $LaTiO_3$/$SrTiO_3$ superlattices, reproduced with permission.[21] Copyright 2015, Springer Nature. d) Thermopower in $SrTiO_3$/$SrTi_{0.8}Nb_{0.2}O_3$ superlattices with varying $SrTi_{0.8}Nb_{0.2}O_3$ thickness, reproduced with permission.[24] Copyright 2007, Springer Nature. e) Polarization hysteresis of Ti-based superlattices, reproduced with permission.[25] Copyright 2007, Wiley-VCH. f) Polarization hysteresis of Ti-based tri-color superlattices, reproduced with permission.[10] Copyright 2005, Springer Nature. g) Observation on polar vortex in $PbTiO_3$/$SrTiO_3$ superlattices using electron microscopy.[27] Copyright 2016, Springer Nature. h) Octahedral tilt angle of different thickness of $SrTiO_3$/$CaTiO_3$ superlattices from ADF and iDPC image, reproduced with permission.[28, 29] Copyright 2022, Springer Nature. i) Temperature dependent muon ratio affected by local magnetic field and normalized permittivity in $LaNiO_3$/$LaAlO_3$ superlattices grown on $LaSrAlO_4$ and $SrTiO_3$ each, reproduced with permission.[34] Copyright 2011, American Association for the Advancement of Science. j) Temperature dependent magnetic hysteresis of $(LuFeO_3)_9/(LuFe_2O_4)_1$ superlattices, reproduced with permission.[35] Copyright 2016, Springer Nature.

**Table 2.** Summary of reported 4*d* $SrRuO_3$ based superlattices.[41, 54, 55, 70, 86-164]

| 4*d* $SrRuO_3$ based superlattices | Functionality of partner compounds | Reference |
|---|---|---|
| $SrRuO_3/SrTiO_3$ | Quantum paraelectric | [54, 55, 70, 86-110] |
| $SrRuO_3/BaTiO_3$ | | [111, 112] |
| $SrRuO_3/PbTiO_3$ | Ferroelectric | [41, 114] |
| $SrRuO_3/PbZr_{0.52}Ti_{0.48}O_3$ | | [113, 115] |
| $SrRuO_3/LaAlO_3$ | Dielectric | [116] |
| $SrRuO_3/SrCuO_2$ | Infinite layer | [117] |
| $SrRuO_3/La_{0.7}Sr_{0.3}MnO_3$ | | [118-137, 143] |
| $SrRuO_3/Pc_{0.7}Ca_{0.3}MnO_3$ | | [138-143] |
| $SrRuO_3/La_{0.7}Ca_{0.3}MnO_3$ | | [144-146] |
| $SrRuO_3/PrMnO_3$ | (Anti-)Ferromagnetic | [147-149] |
| $SrRuO_3/SrMnO_3$ | | [150-155] |
| $SrRuO_3/LaCoO_3$ | | [156] |
| $SrRuO_3/LaNiO_3$ | | [157] |
| $SrRuO_3/BiFeO_3$ | Multiferroic | [158-160] |
| $SrRuO_3/SrIrO_3$ | | [161-163] |
| $SrRuO_3/SrIrO_3/SrZrO_3$<br>$SrRuO_3/SrHfO_3/SrZrO_3$ | 5*d* oxides | [164] |

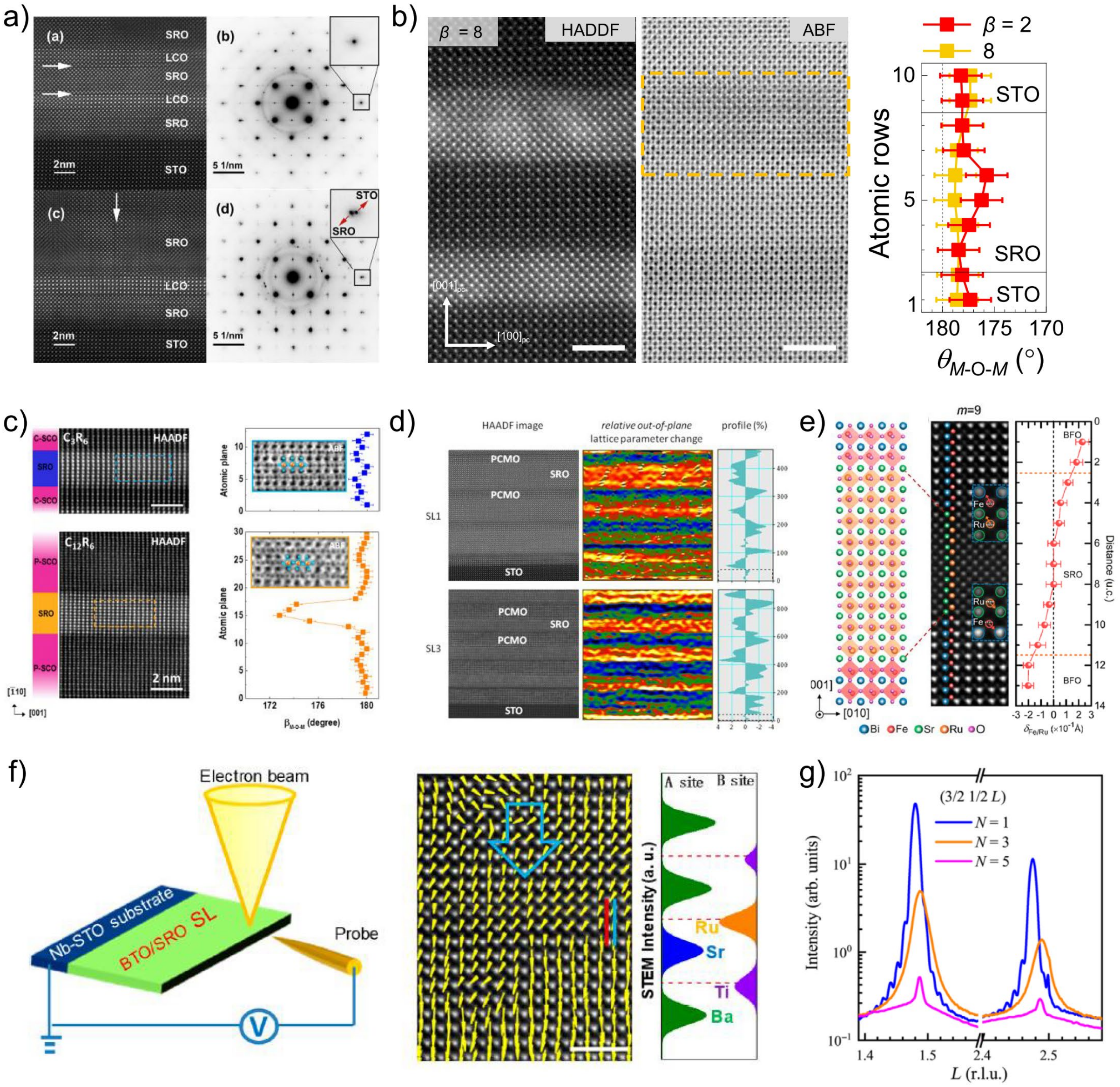

**Figure 3.** a) Cross-sectional scanning transmission electron microscopy and selected area electron diffraction images of $SrRuO_3/LaCoO_3$ multilayer thin films, reproduced with permission.[156] Copyright 2018, Elsevier. b) Scanning transmission electron microscopy images for $SrRuO_3/SrTiO_3$ superlattices and propagation of octahedral distortion with respect to $SrTiO_3$ layer thickness, reproduced with permission.[91] Copyright 2020, Wiley-VCH. c) Scanning transmission electron microscopy image and octahedral distortion of $SrRuO_3/SrCuO_2$ superlattices, reproduced with permission.[117] Copyright 2021, American Chemical Society. d) Scanning transmission electron microscopy image and relative out-of-plane lattice parameter change of $SrRuO_3/Pr_{0.7}Ca_{0.3}MnO_3$ superlattices, reproduced with permission.[141] Copyright 2011, IOP publishing. e) Ionic displacements of interfacial $SrRuO_3$ induced by ferroelectric polarization of $BiFeO_3$ in $SrRuO_3/BiFeO_3$ multilayer structure, reproduced with permission.[158] Copyright 2022, American Chemical Society. f) In situ

transmission electron microscopy for electrical field control of Ru ion displacement through ferroelectric polarization in $SrRuO_3/BaTiO_3$ superlattices, reproduced with permission.[112] Copyright 2020, American Chemical Society. g) Observation on octahedral distortion of $SrRuO_3/SrTiO_3$ superlattice by half-order X-ray diffraction measurements, reproduced with permission.[107] Copyright 2020, American Association for the Advancement of Science.

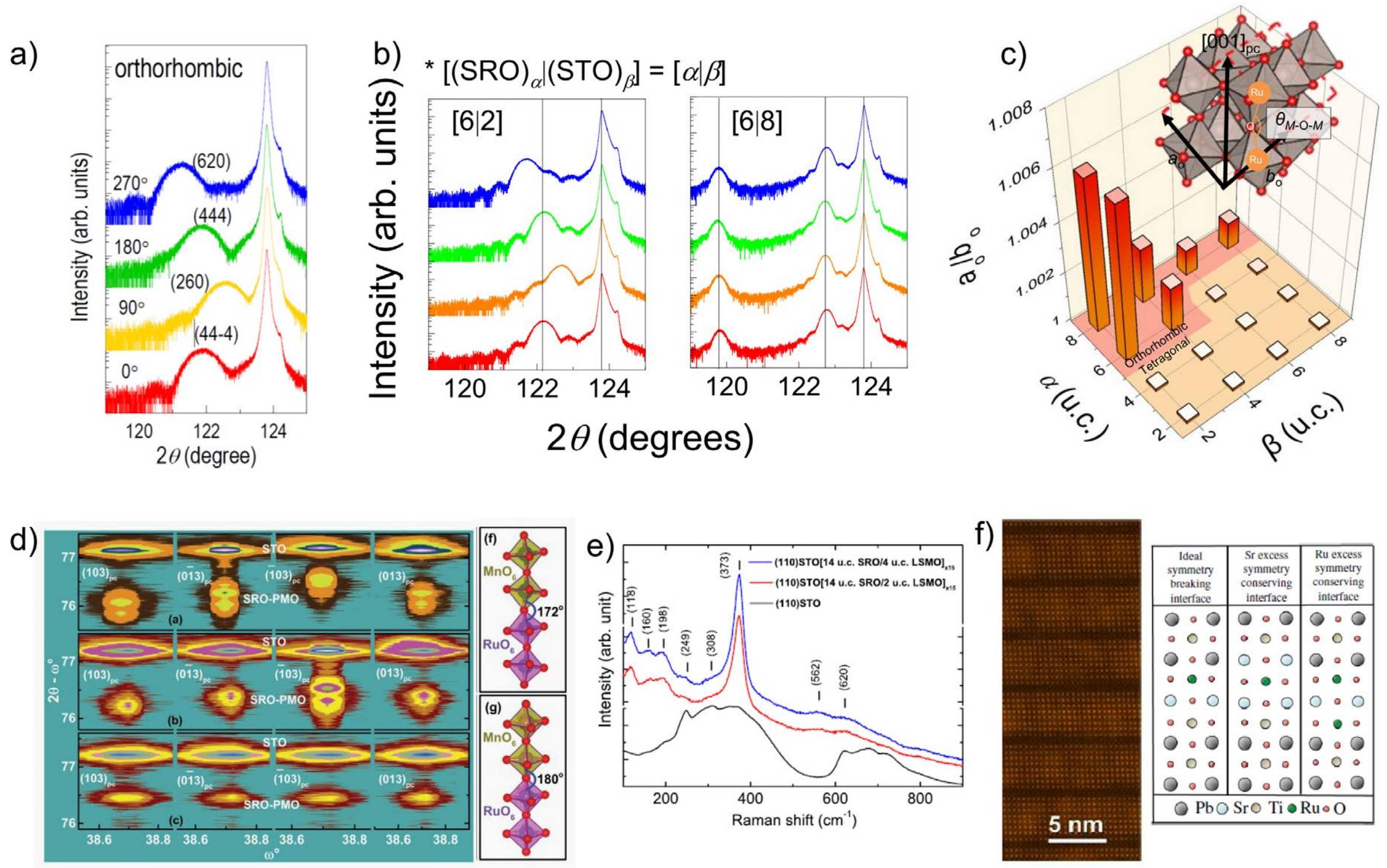

**Figure 4.** a) Asymmetric crystal structure of orthorhombic $SrRuO_3$ heterostructure observed by off-axis measurement around (204) $SrTiO_3$ Bragg reflections with four different azimuth angles, reproduced with permission.[49] Copyright 2017, Springer Nature. b) Structural phase transition of $SrRuO_3/SrTiO_3$ superlattices from orthorhombic to tetragonal as $SrTiO_3$ layer thickness increases.[91] Copyright 2020, Wiley-VCH. c) Octahedral distortion mapping of $SrRuO_3/SrTiO_3$ superlattices with respect to $SrRuO_3$ and $SrTiO_3$ layer thickness, reproduced with permission.[91] Copyright 2020, Wiley-VCH. d) Structural phase transition of $SrRuO_3/PrMnO_3$ with respect to $PrMnO_3$ layer thickness shown by reciprocal space mapping around (103) $SrTiO_3$ Bragg reflections for four different azimuth angles, reproduced with permission.[147] Copyright 2018, Wiley-VCH. e) Raman spectroscopy of (110) oriented $SrTiO_3$ and $SrRuO_3/La_{0.7}Sr_{0.3}MnO_3$ superlattice on (110) oriented $SrTiO_3$ with different $La_{0.7}Sr_{0.3}MnO_3$ thickness, reproduced with permission.[129] Copyright 2014, AIP publishing. f) Scanning transmission electron microscopy image and the three types of theoretically possible

interfaces of an $SrRuO_3/PbTiO_3$ superlattice, reproduced with permission.[41] Copyright 2012, American Physical Society.

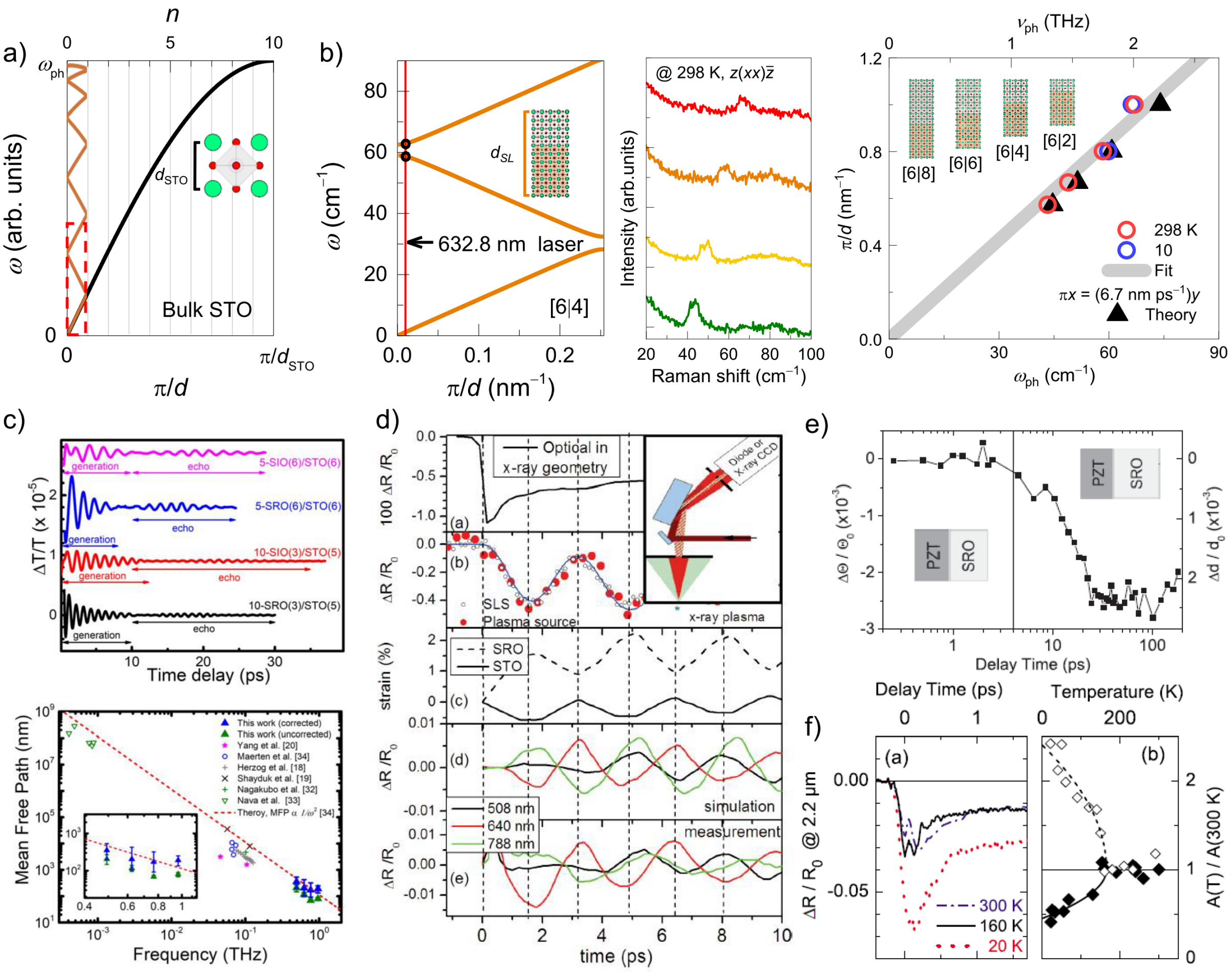

**Figure 5.** a) Schematic of zone-folded phonon dispersion of bulk $SrTiO_3$ and $SrRuO_3/SrTiO_3$ superlattice, reproduced with permission.[88] Copyright 2022, Wiley-VCH. b) Theoretical estimation of zone-folded phonon dispersion of $SrRuO_3/SrTiO_3$ superlattice system and Raman spectra. reproduced with permission.[88] Copyright 2022, Wiley-VCH. c) Time-resolved pump-probe of $SrRuO_3$(or $SrIrO_3$)/$SrTiO_3$ superlattice and frequency dependent phonon mean free path, reproduced with permission.[103] Copyright 2020, IOP publishing. d) Comparison of zone-folded acoustic phonon in $SrRuO_3/SrTiO_3$ superlattice observed by optical pump-probe measurement and ultrafast X-ray diffraction, reproduced with permission.[108] Copyright 2012, American Physical Society. e) relative angular shift of $SrRuO_3/PbZr_{0.2}Ti_{0.8}O_3$ of the selected Bragg reflection, reproduced with permission.[113] Copyright 2007, American Physical Society. f) Optical reflectivity change for three different temperatures and comparison between temperature dependence of the amplitude of the reflectivity change and superlattice phonon oscillation, reproduced with permission.[95] Copyright 2008, American Physical Society.

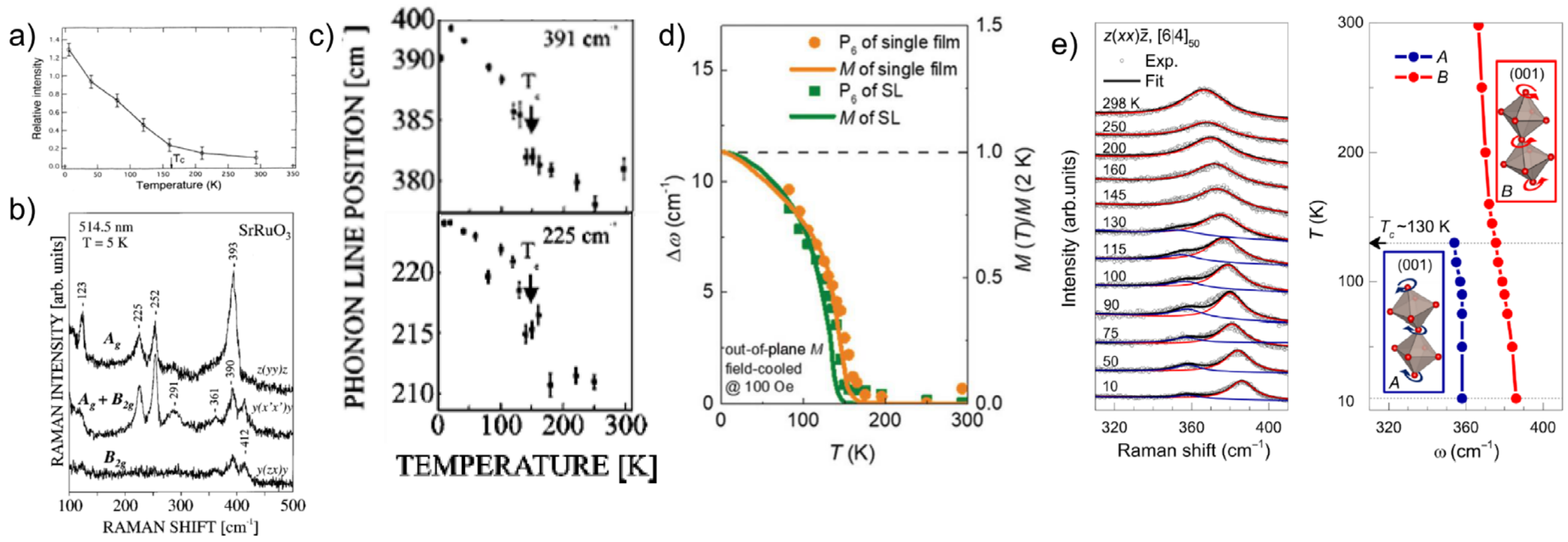

**Figure 6.** a) Temperature-dependent Raman spectra of $SrRuO_3$ thin film showing phonon anomalies at ferromagnetic transition temperature, reproduced with permission.[173] Copyright 1995, American Physical Society. b) Polarized Raman spectra of ~300-nm-thickness $SrRuO_3$ thin film on (010) and (001) $SrTiO_3$ substrates, reproduced with permission.[174] Copyright 1999, American Physical Society. c) Temperature dependence of the Raman peak position of (010)-oriented $SrRuO_3$ thin film, reproduced with permission.[174] Copyright 1999, American Physical Society. d) Temperature dependence of both phonon anomaly and magnetization to compare spin-phonon coupling of single $SrRuO_3$ film and $SrRuO_3/SrTiO_3$ superlattice, reproduced with permission.[70] Copyright 2020, The Royal Society of Chemistry. e) Phonon mode splitting corresponding to oxygen vibrations with orthogonal polarizations in bulk $SrRuO_3$ below ferromagnetic phase transition temperature, reproduced with permission.[55] Copyright 2022, American Association for the Advancement of Science.

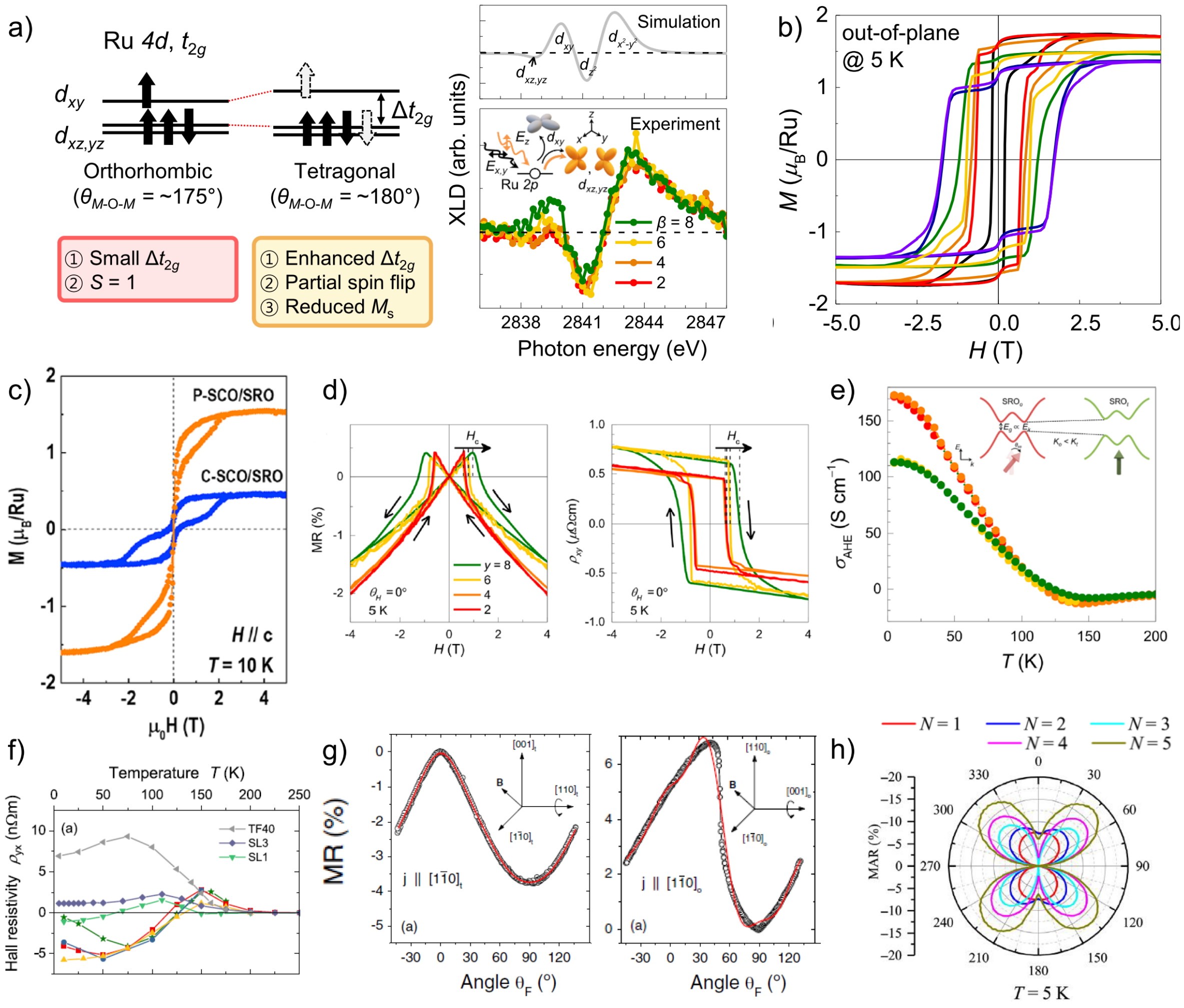

**Figure 7.** a) Orbital energy splitting difference between orthorhombic and tetragonal $SrRuO_3$, reproduced with permission.[91] Copyright 2020, Wiley-VCH. b) Systematic change of saturation magnetization with structural phase transition of $SrRuO_3/SrTiO_3$ superlattice between orthorhombic and tetragonal, reproduced with permission.[91] Copyright 2020, Wiley-VCH. c) Structural dependence of magnetic field-dependent magnetization of $SrCuO_2/SrRuO_3/SrCuO_2$ heterostructure with respect to the $SrCuO_2$ layer thickness, reproduced with permission.[117] Copyright 2021, American Chemical Society. d) Magnetoresistance and anomalous Hall effect in $SrRuO_3/SrTiO_3$ superlattices, reproduced with permission.[89] Copyright 2021, American Chemical Society. e) Temperature dependent anomalous Hall conductivity of $SrRuO_3$ in orthorhombic and tetragonal structure, reproduced with permission.[89] Copyright 2021, American Chemical Society. f) Temperature dependent Hall resistivity of $SrRuO_3/Pr_{0.7}Ca_{0.3}MnO_3$ superlattice, reproduced with permission.[141] Copyright 2011, IOP publishing. g) Magnetic field angle dependent magnetoresistance measurement of orthorhombic and tetragonal $SrRuO_3/Pr_{0.7}Ca_{0.3}MnO_3$ superlattices, reproduced with permission.[138] Copyright 2019, IOP publishing. h) Magnetic field angle

dependent magnetoresistance of $SrRuO_3/SrTiO_3$ superlattices with different $SrTiO_3$ thickness, reproduced with permission.[107] Copyright 2020, American Association for the Advancement of Science.

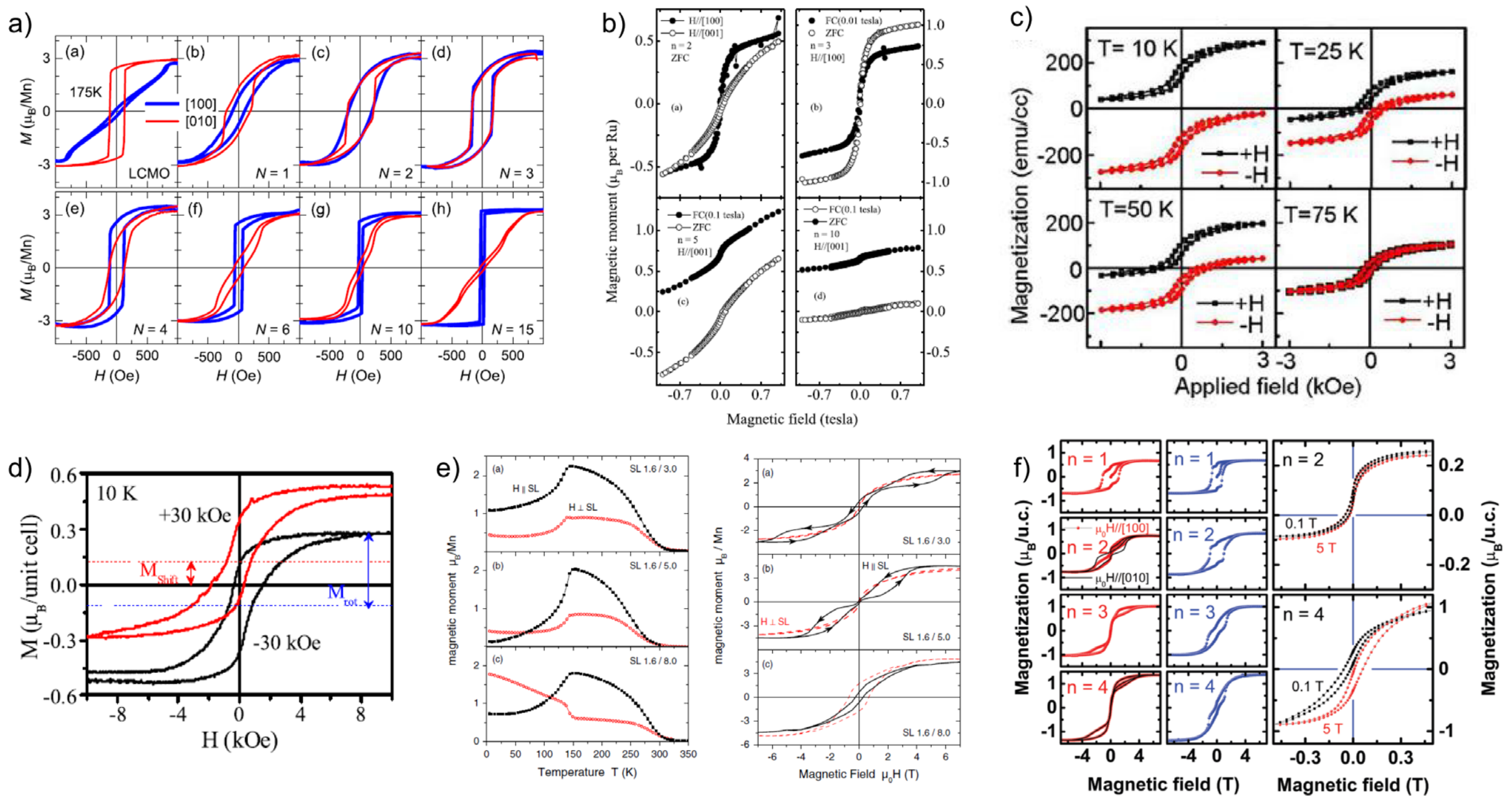

**Figure 8.** a) Repetition number dependent lateral magnetic anisotropy of $SrRuO_3/La_{0.67}Ca_{0.33}MnO_3$ superlattices, reproduced with permission.[145] Copyright 2021, AIP publishing. b) Magnetic hysteresis loop showing dependence in field direction and difference between field-cooled and zero-field-cooled hysteresis of $SrRuO_3/SrMnO_3$ superlattices, reproduced with permission.[151] Copyright 2005, Springer Nature. c) Temperature dependence of exchange bias effect in $SrRuO_3/LaNiO_3$ superlattices, reproduced with permission.[157] Copyright 2020, Taylor & Francis. d) Exchange bias effect in $SrRuO_3/BiFeO_3$ superlattices, reproduced with permission.[159] Copyright 2019, Elsevier. e) Left panel, temperature dependence of out-of-plane (circles) and in-plane (squares) magnetic moment measured by field cooled method. Right panel, $SrRuO_3$ thickness-dependent magnetic moment in $SrRuO_3/La_{0.7}Sr_{0.3}MnO_3$ superlattice, reproduced with permission.[136] Copyright 2010, American Physical Society. f) In-plane, out-of-plane, and field-cooled in-plane magnetic hysteresis loop of $SrRuO_3/PrMnO_3$ superlattice as a function of $PrMnO_3$ thickness, reproduced with permission.[147] Copyright 2018, Wiley-VCH.

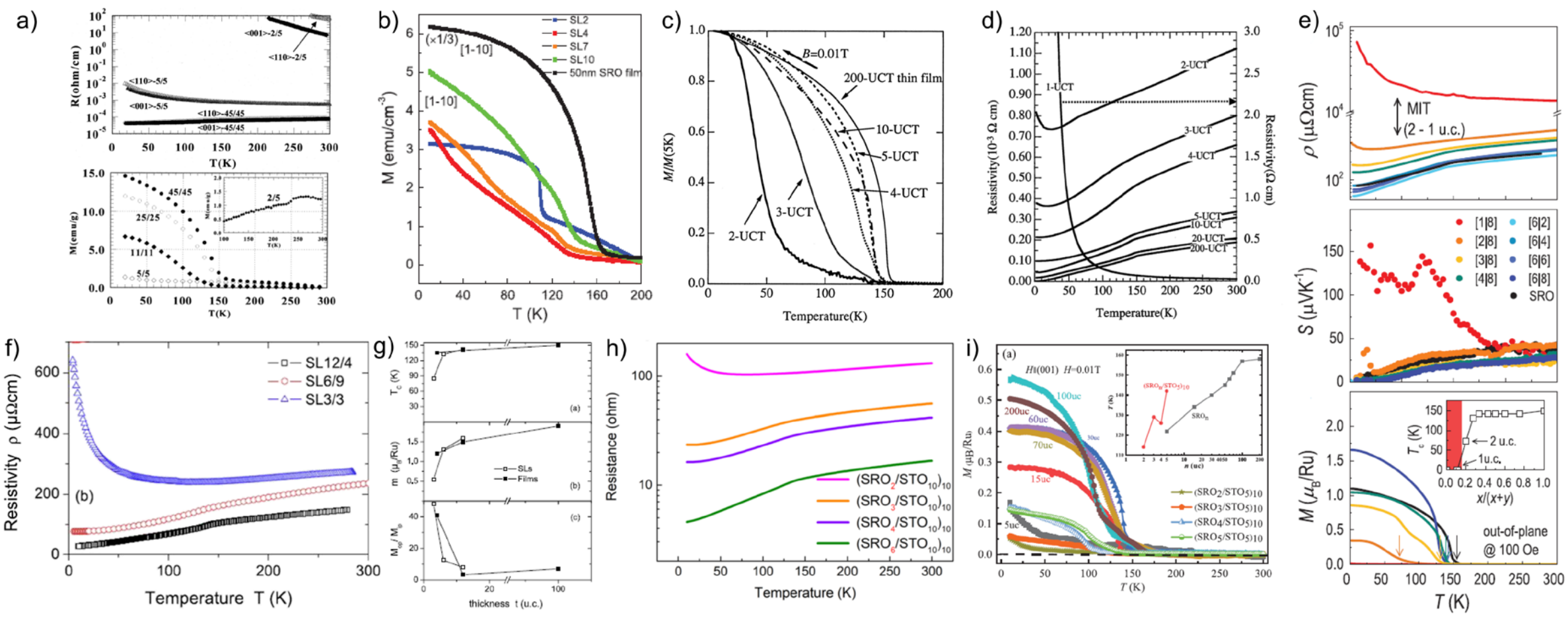

**Figure 9.** a) Temperature dependent transport and magnetization with respect to stacking periodicity in of $SrRuO_3/BaTiO_3$ superlattice grown on $SrTiO_3$ (110) substrates, reproduced with permission.[111] Copyright 2000, Elsevier. b) Temperature dependent magnetization of $SrRuO_3/LaAlO_3$ superlattices with varying $SrRuO_3$ layer thickness, reproduced with permission.[116] Copyright 2012, AIP publishing. c) Temperature dependent magnetization of $SrRuO_3/SrTiO_3$ superlattices with varying $SrRuO_3$ thickness, reproduced with permission.[86, 87] Copyright 1998, Elsevier. d) Temperature dependent transport of $SrRuO_3/SrTiO_3$ superlattices with different $SrRuO_3$ thickness, reproduced with permission.[86, 87] Copyright 1998, The Physical Society of Japan e) Dimensional crossover effect of $SrRuO_3/SrTiO_3$ superlattice as a function of $SrRuO_3$ thickness, reproduced with permission[92] Copyright 2020, American Physical Society. f) Temperature dependent resistivity and g) summarized magnetic properties of three different $SrRuO_3/SrTiO_3$ superlattices, reproduced with permission.[100] Copyright 2013, IOP publishing. h) Temperature dependent resistance of $SrRuO_3/SrTiO_3$ superlattices with varying $SrRuO_3$ layer thickness, reproduced with permission.[98] Copyright 2020, American Physical Society. i) Temperature dependent magnetization $SrRuO_3/SrTiO_3$ superlattices as a function of $SrRuO_3$ thickness, reproduced with permission.[54] Copyright 2022, Wiley-VCH.

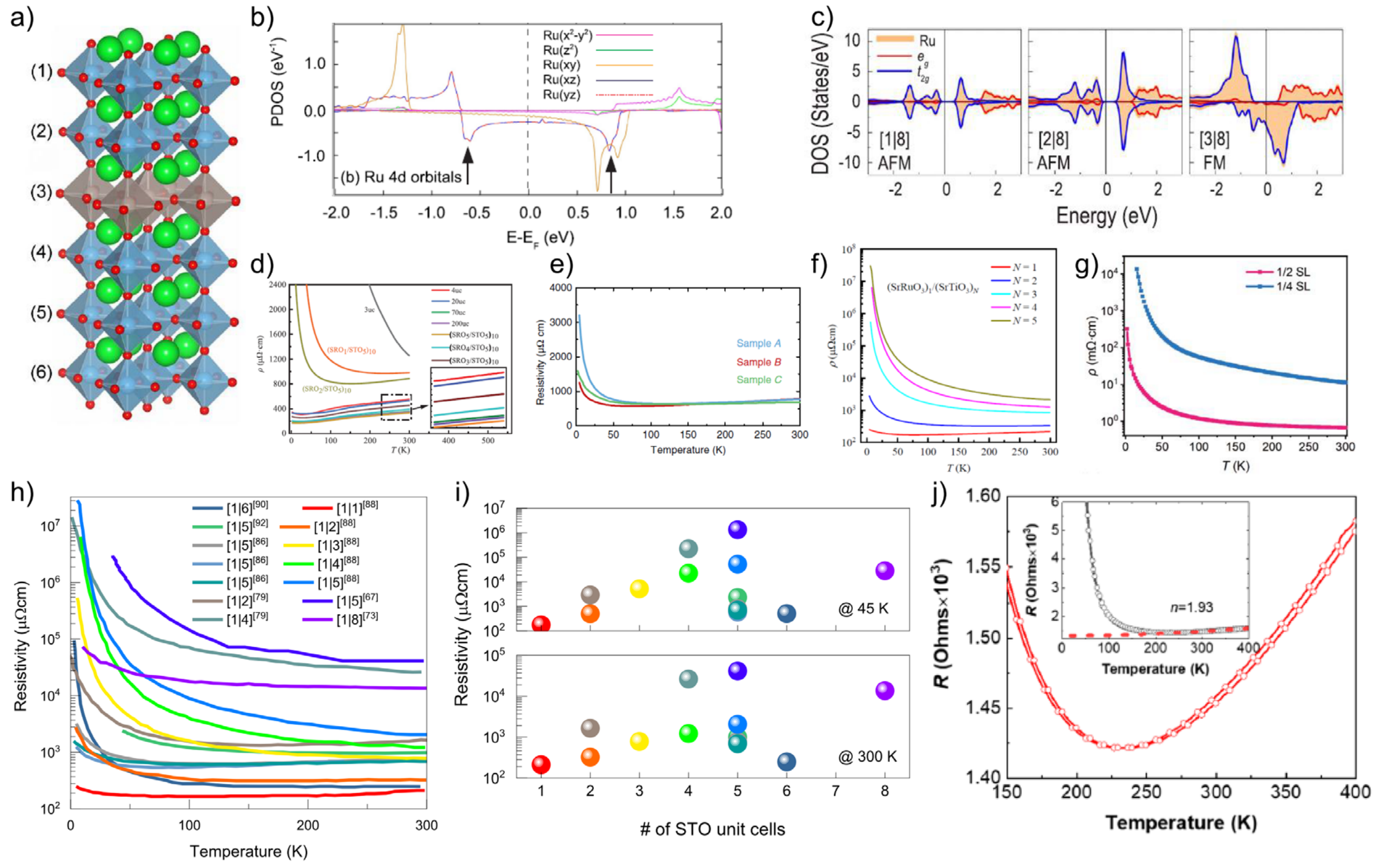

**Figure 10.** a) Schematic and b) projected density of states of Ru orbitals of oxide superlattice composed monolayer $SrRuO_3$ layer and 5 unit cell layers of $SrTiO_3$, reproduced with permission.[102] Copyright 2012, American Physical Society. c) Dimensionality-induced phase transition of $SrRuO_3/SrTiO_3$ superlattice as a function of the thickness of the $SrRuO_3$ layer, reproduced with permission.[92] Copyright 2020, American Physical Society. d) Temperature dependent transport with varying $SrRuO_3$ layer thickness in $SrRuO_3/SrTiO_3$ superlattice, reproduced with permission.[178] Copyright 2020, Wiley-VCH. e) Temperature dependent transport of three different monolayer $SrRuO_3$ superlattices with the same periodicity, reproduced with permission.[105] Copyright 2019, American Physical Society. f) Temperature dependent resistivity of $SrRuO_3/SrTiO_3$ superlattice with varying $SrTiO_3$ thickness, reproduced with permission.[107] Copyright 2020, American Association for the Advancement of Science. g) Temperature dependent transport of $SrRuO_3/SrTiO_3$ superlattice with two different $SrTiO_3$ thickness, reproduced with permission.[99] Copyright 2022, Springer Nature. h) Summary of the temperature dependent resistivity of $SrRuO_3$ monolayer within $SrRuO_3/SrTiO_3$ superlattices.[54, 86, 92, 99, 105, 107, 109] i) $SrTiO_3$ thickness dependent resistivity of monolayer $SrRuO_3$ superlattice at 45 K and 300 K. j) Temperature dependent transport of monolayer $SrRuO_3$ superlattice with 10 unit cell layers of $BaTiO_3$, reproduced with permission.[112] Copyright 2021, American Chemical Society.

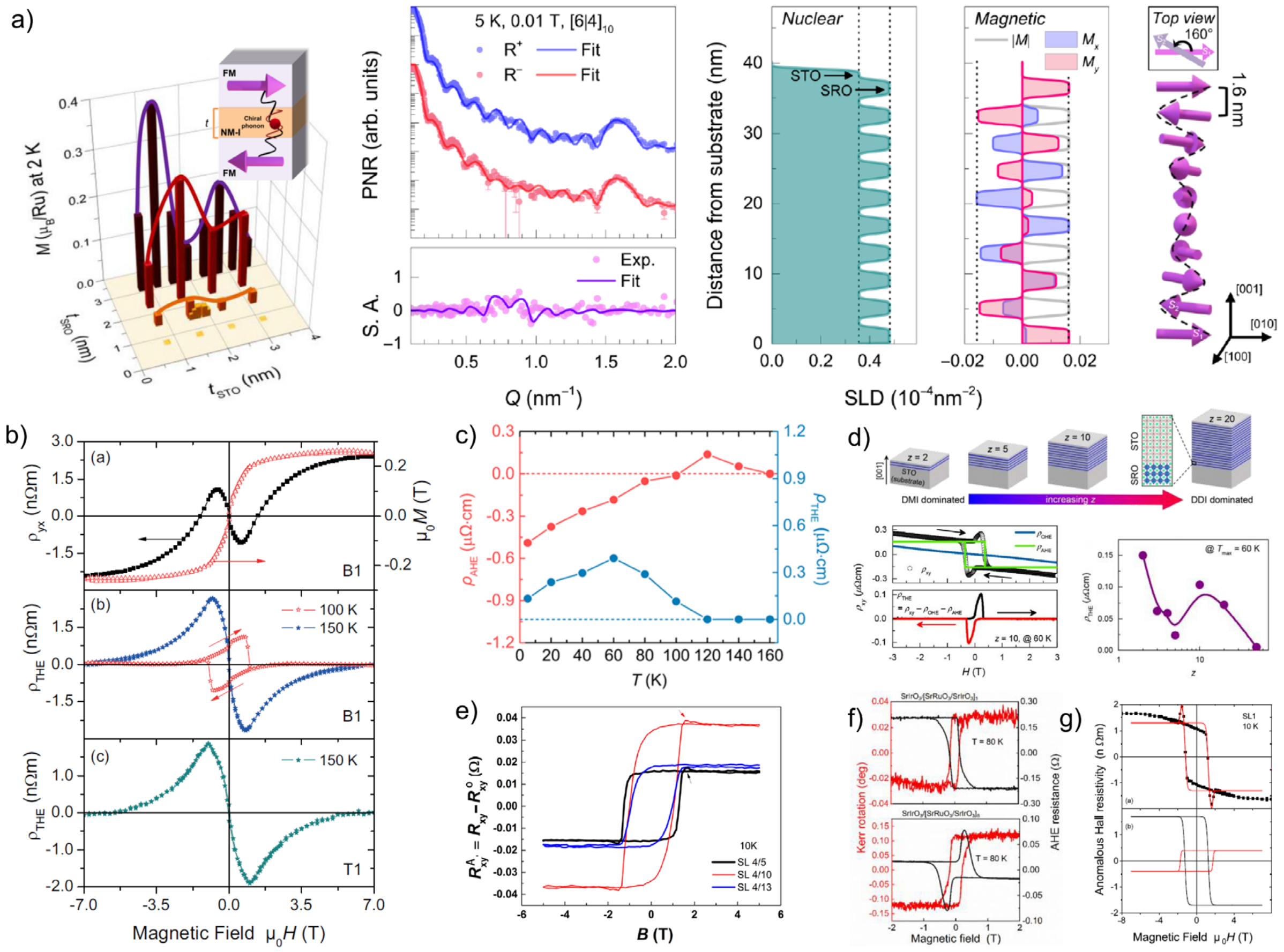

**Figure 11.** a) In-plane magnetization in $SrRuO_3/SrTiO_3$ superlattices at 2 K as a function of $SrRuO_3$ and $SrTiO_3$ layer thickness. The middle and right panels are polarized neutron reflectivity result with fitting analysis and corresponding spin structure of $SrRuO_3/SrTiO_3$ superlattices, respectively, reproduced with permission.[55] Copyright 2022, American Association for the Advancement of Science. b) Interlayer coupling induced topological Hall effect in $SrRuO_3/La_{0.7}Sr_{0.3}MnO_3$ superlattice, reproduced with permission.[122] Copyright 2017, Wiley-VCH. c) Temperature dependent anomalous and topological Hall effect in $SrRuO_3/BiFeO_3$ multilayer, reproduced with permission.[158] Copyright 2022, American Chemical Society. d) Repetition number dependence of topological Hall effect in $SrRuO_3/SrTiO_3$ superlattice, reproduced with permission.[90] Copyright 2021, Elsevier. e) Topological Hall signal of $SrRuO_3/SrIrO_3$ superlattice, reproduced with permission.[163] Copyright 2021, American Chemical Society. f) Kerr rotation and anomalous Hall effect with hump structure formed as $SrIrO_3/SrRuO_3/SrIrO_3$ trilayers changed to $SrRuO_3/SrIrO_3$ superlattice, reproduced with permission.[162] Copyright 2021, American Physical Society. g) Anomalous Hall resistivity with hump signal in $SrRuO_3/Pr_{0.7}Ca_{0.3}MnO_3$ superlattice, reproduced with permission.[138] Copyright 2019, IOP publishing.

**Table 3.** Summary of reported 5*d* $SrIrO_3$ based superlattices.[75, 199-218]

| 5*d* $SrIrO_3$ based superlattices | Functionality of partner compounds | Reference |
|---|---|---|
| $SrIrO_3/SrTiO_3$ | Quantum paraelectric | [75, 199-205] |
| $SrIrO_3/CaTiO_3$ | Incipient ferroelectric | [206] |
| $SrIrO_3/La_{1-x}Sr_xMnO_3$ | (Anti-)Ferromagnetic | [207-211] |
| $SrIrO_3/SrMnO_3$ | | [212] |
| $SrIrO_3/LaCoO_3$ | | [213, 214] |
| $SrIrO_3/LaFeO_3$ | | [215] |
| $SrIrO_3/LaMnO_3$ | | [216] |
| $SrIrO_3/LaNiO_3$ | | [217] |
| $SrIrO_3/CaMnO_3$ | | [218] |

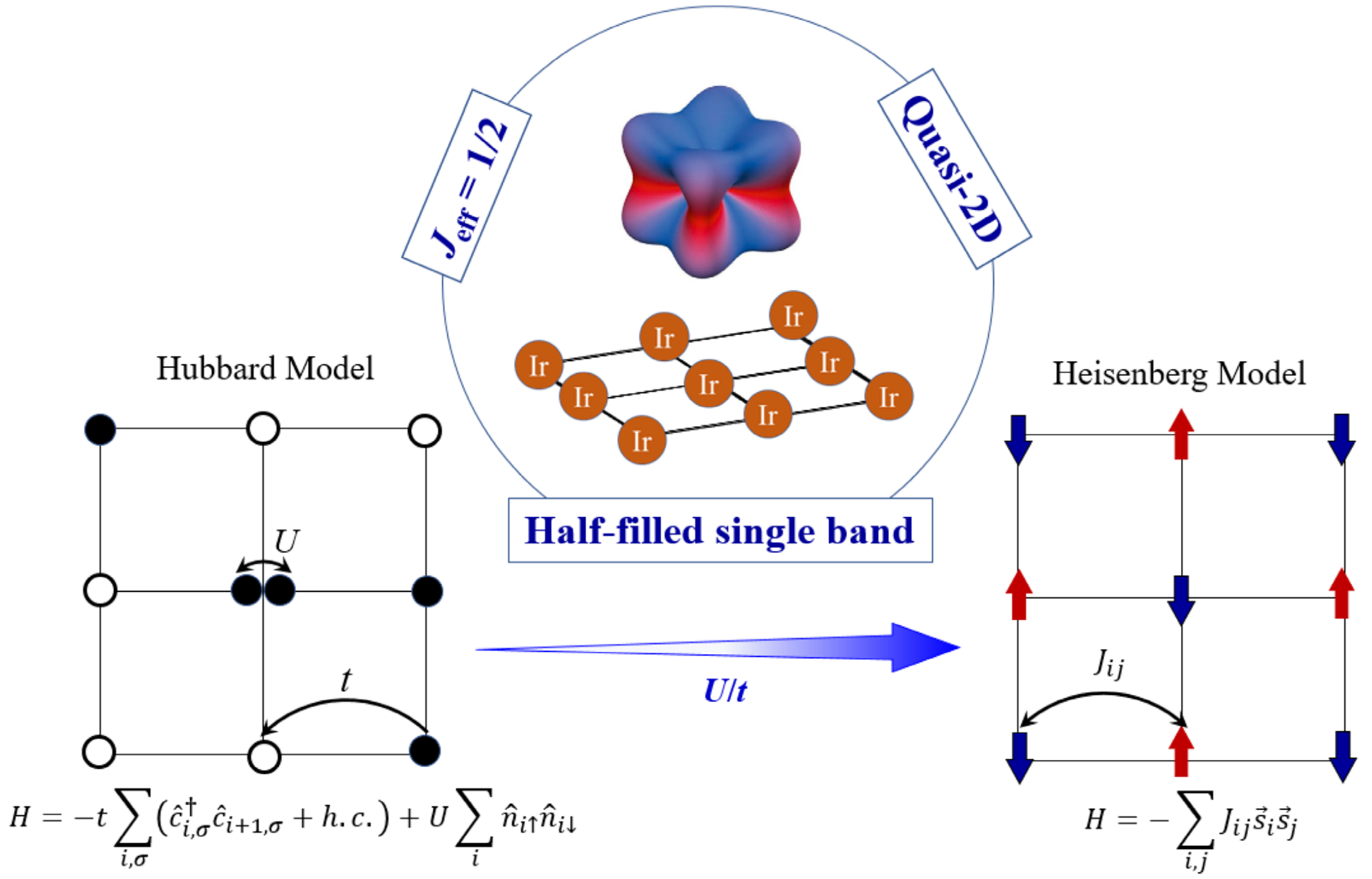

**Figure 12.** Physics relevant to the 2D square-lattice iridate, ranging from the charge Hubbard model to the spin Heisenberg model.

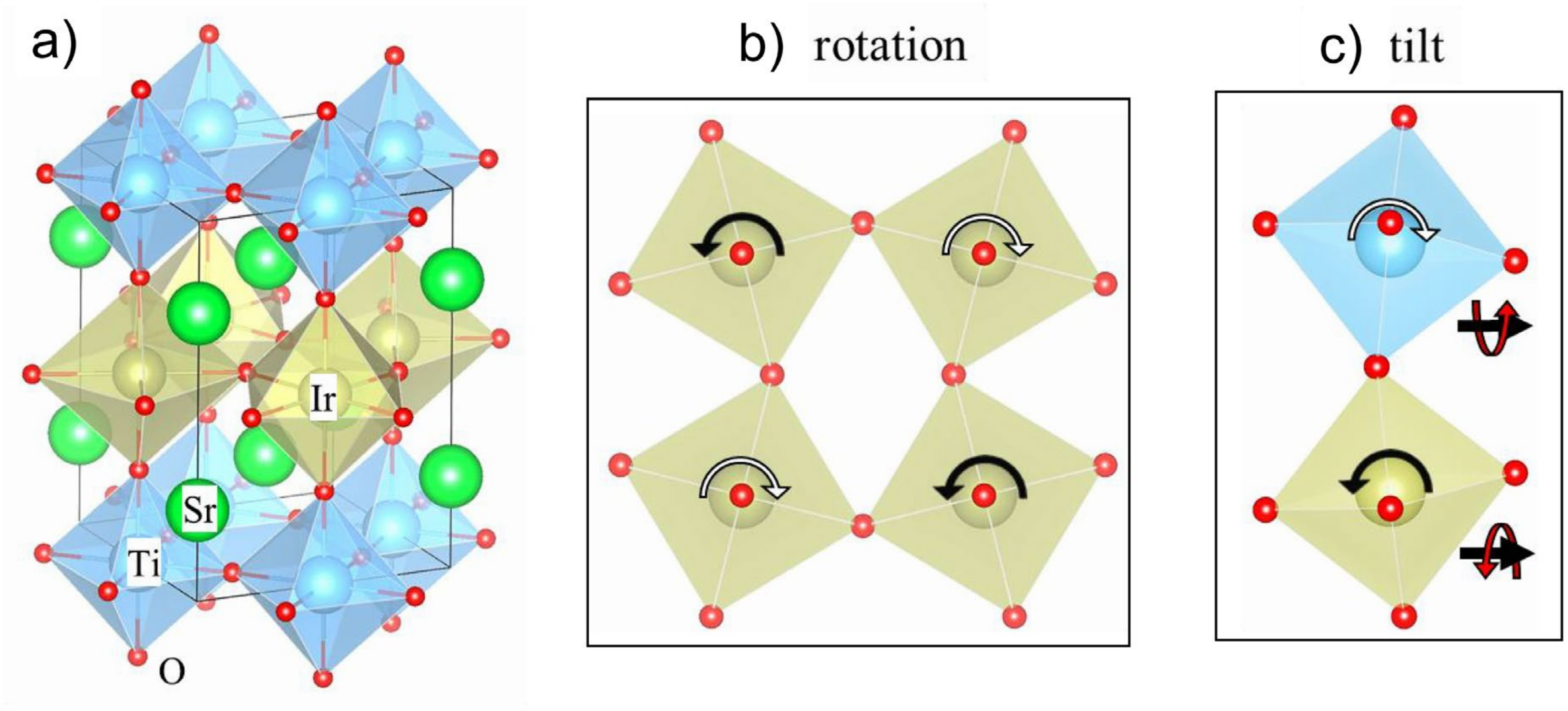

**Figure 13.** a) Lattice structure of $(SrIrO_3)1/(SrTiO_3)1$. Schematics of b) octahedral rotation and c) tilt.

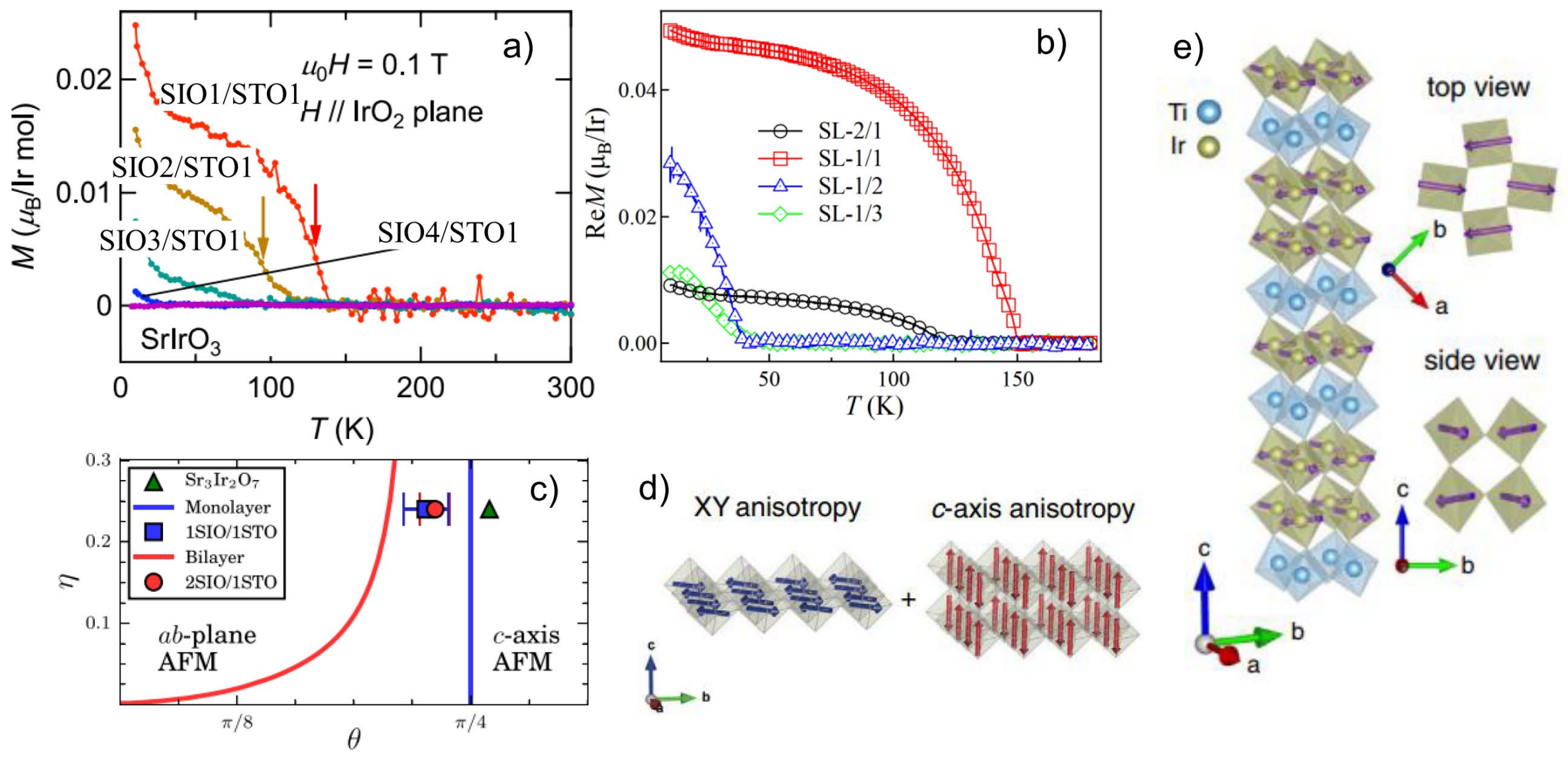

**Figure 14.** a) Temperature dependent magnetization of $(SrIrO_3)n/(SrTiO_3)1$, reproduced with permission.[75] Copyright 2015, American Physical Society. b) Temperature dependent magnetization of $(SrIrO_3)1/(SrTiO_3)m$, reproduced with permission.[199] Copyright 2017, American Physical Society. c) Classic magnetic phase diagram of monolayer- and bilayer-iridates. (here, $\eta$ is the ratio of Hund's coupling on Coulomb repulsion and θ denotes tetragonal distortion of $IrO_6$ octahedron), reproduced with permission.[203] Copyright 2019, Springer Nature. d) Schematic of the addition of an easy-plane monolayer iridate and easy-axis bilayer iridate. e) Magnetic structure of a hybrid $(SrIrO_3)1/(SrTiO_3)1/(SrIrO_3)2/(SrTiO_3)1$ superlattice, reproduced with permission.[204] Copyright 2022, American Physical Society.

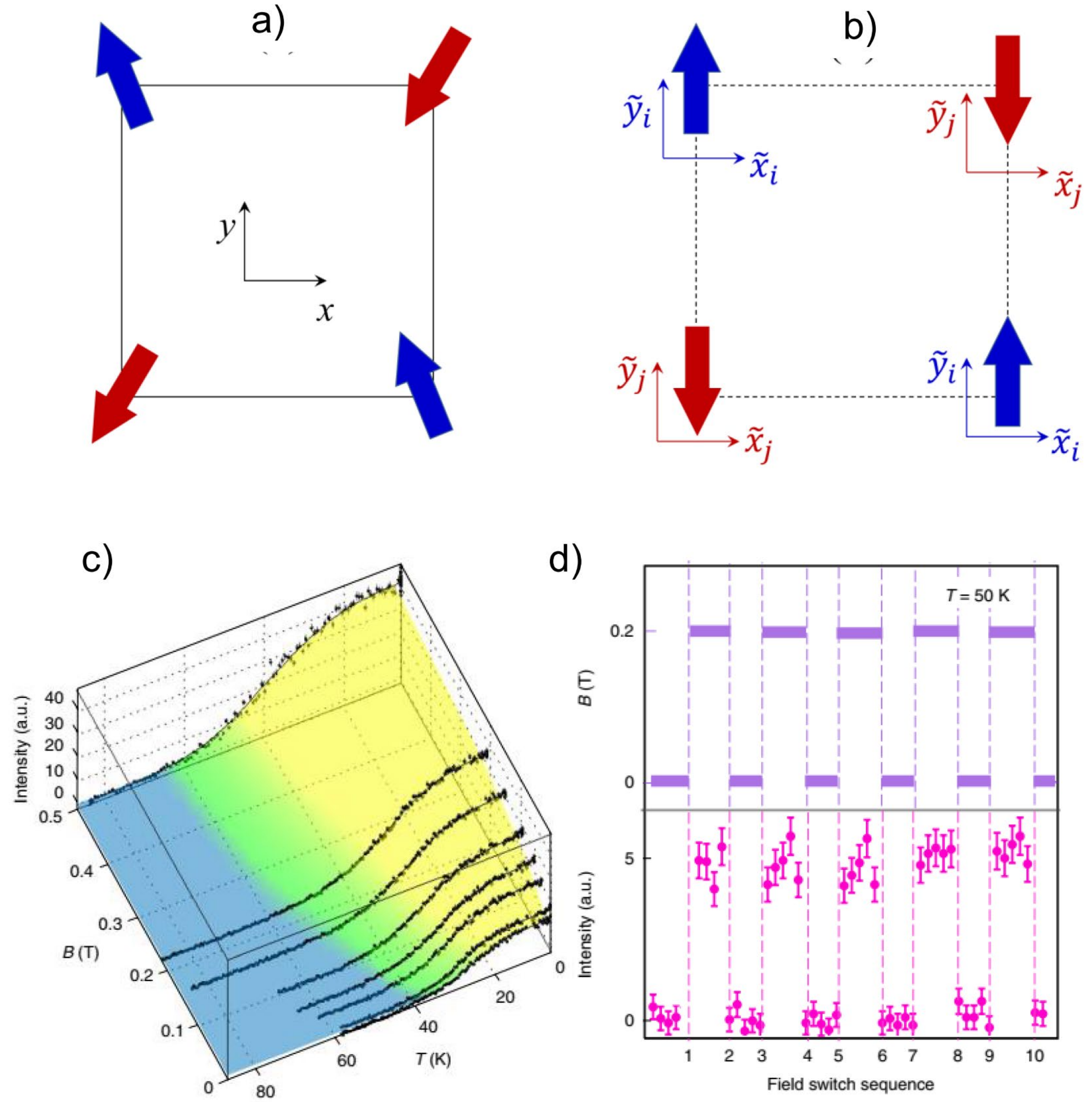

**Figure 15.** Schematics of canted antiferromagnetic structure in a) the global frame and b) a staggered local frame. c) Temperature dependent magnetic peak intensity of $(SrIrO_3)1/(SrTiO_3)2$ under various magnetic fields, reproduced with permission. Copyright 2018, Springer Nature. d) Switching sequence of a 0.2 T magnetic field and the associated change in the magnetic peak intensity of $(SrIrO_3)1/(SrTiO_3)2$, reproduced with permission.[205] Copyright 2018, Springer Nature.

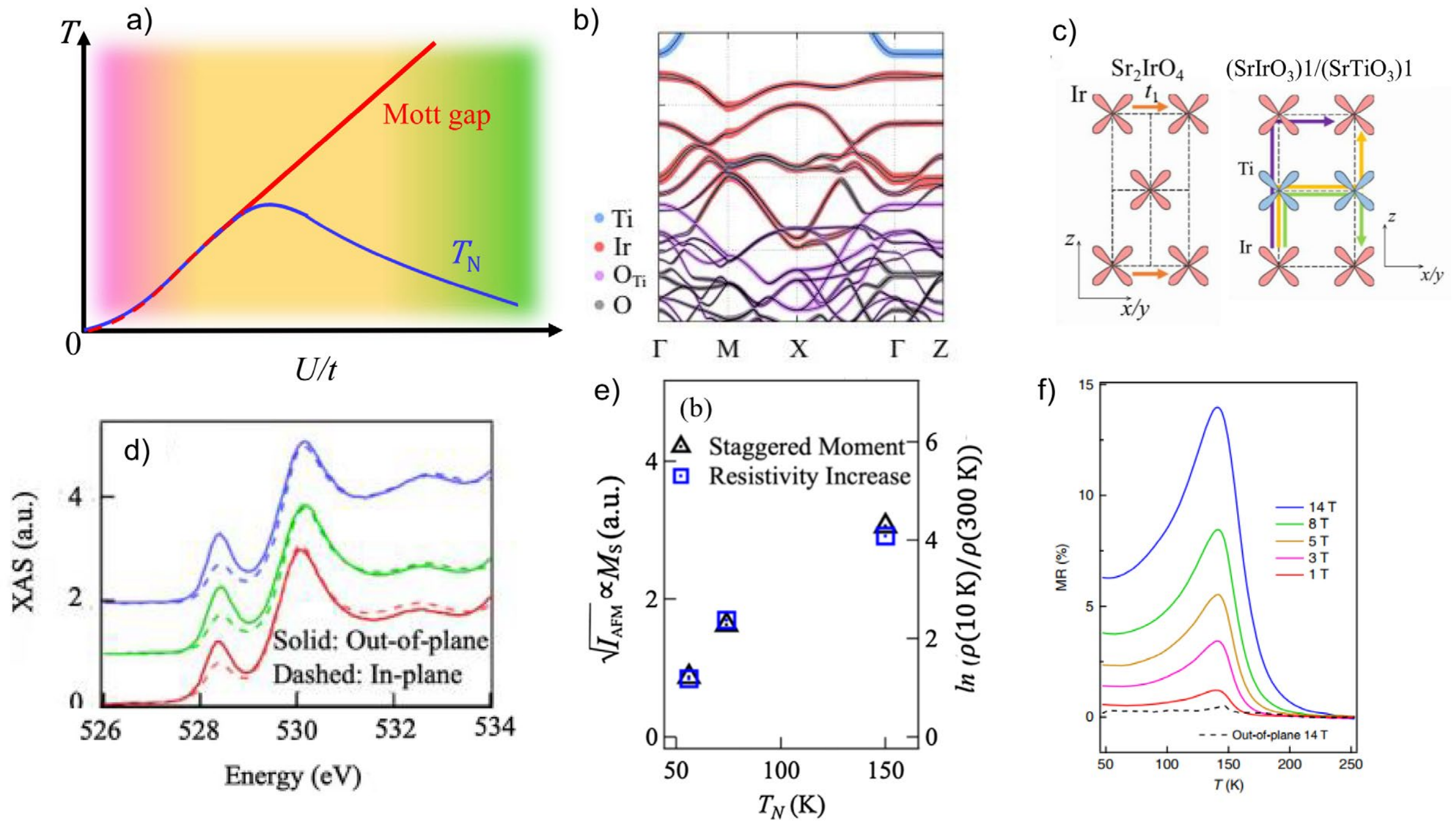

**Figure 16.** a) A generic phase diagram of the half-filled single-band Hubbard system as a function of $U$/t. b) Representative band structure of $(SrIrO_3)1/(SrTiO_3)1$, reproduced with permission.[202] Copyright 2016, American Physical Society. c) Orbital arrangements of Ir and Ti orbitals for $Sr_2IrO_4$ and $(SrIrO_3)1/(SrTiO_3)1$, reproduced with permission.[202] Copyright 2016, American Physical Society. d) Polarization-dependent X-ray absorption spectra of $(SrIrO_3)1/(SrTiO_3)1$ grown on $SrTiO_3$ (red), LSAT (green), and $NdGaO_3$ (blue) substrates, reproduced with permission.[248] Copyright 2020, American Physical Society. e) Staggered moment and resistivity increase of $(SrIrO_3)1/(SrTiO_3)1$ as a function of Neel temperature, reproduced with permission.[248] Copyright 2020, American Physical Society. f) Magnetoresistance dependence on temperature under various magnetic fields, reproduced with permission.[253] Copyright 2019, Springer Nature.

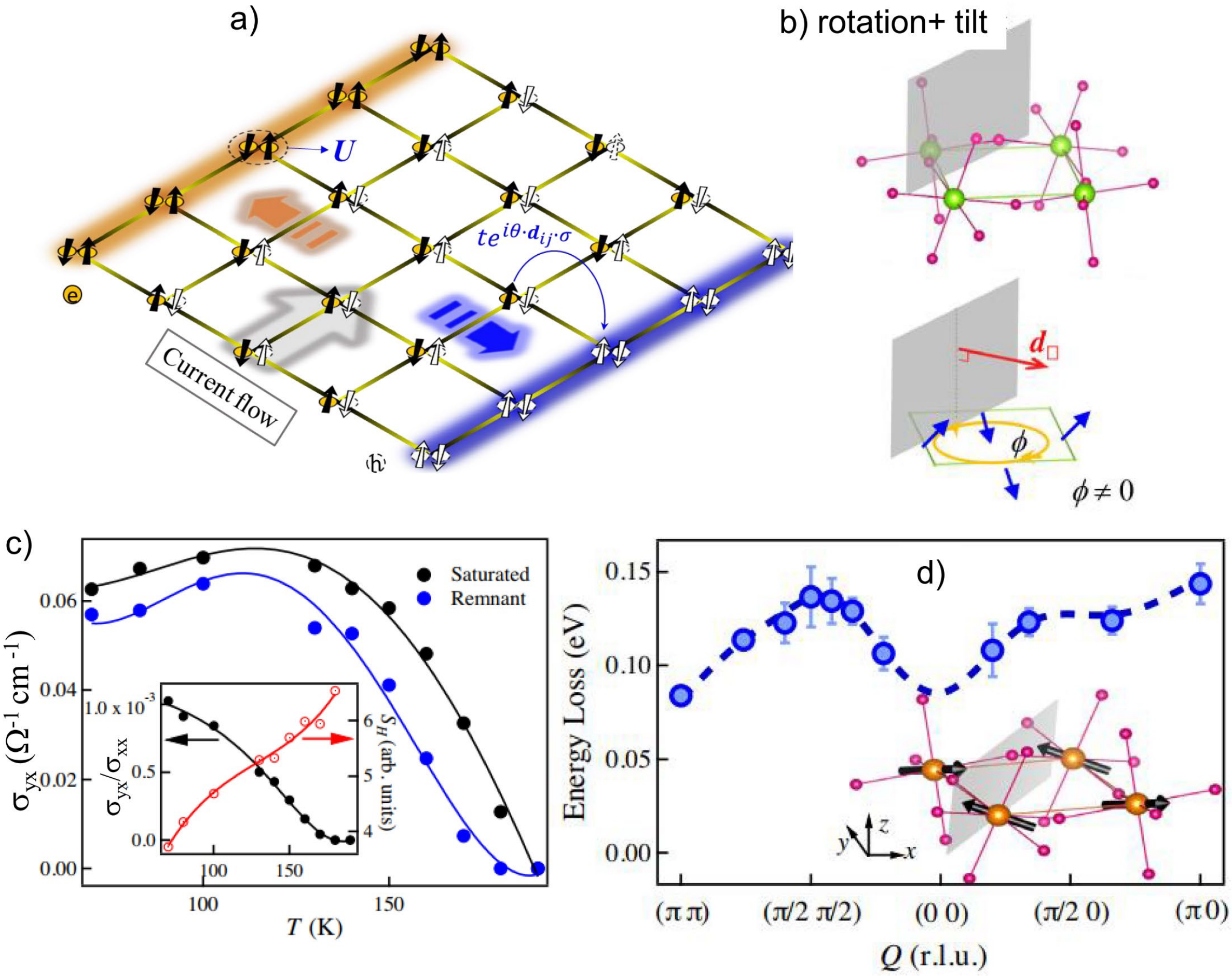

**Figure 17.** a) Schematic of a distorted 2D Hubbard system with a SU(2) gauge field modulating the electron hopping process. b) Square lattice with a finite octahedral rotation and tilt. The bottom panel displays the orientation of the Dzyaloshinskii–Moriya vectors. c) Temperature dependent anomalous Hall conductivity of ($SrIrO_3$)1/($CaTiO_3$)1. The inset displays the temperature dependencies of the anomalous Hall angle (left) and anomalous Hall coefficient (right). d) Magnon dispersion of ($SrIrO_3$)1/($CaTiO_3$)1 at base temperature. The inset displays the ground-state magnetic structure of the superlattice, reproduced with permission.[206] Copyright 2022, American Physical Society.

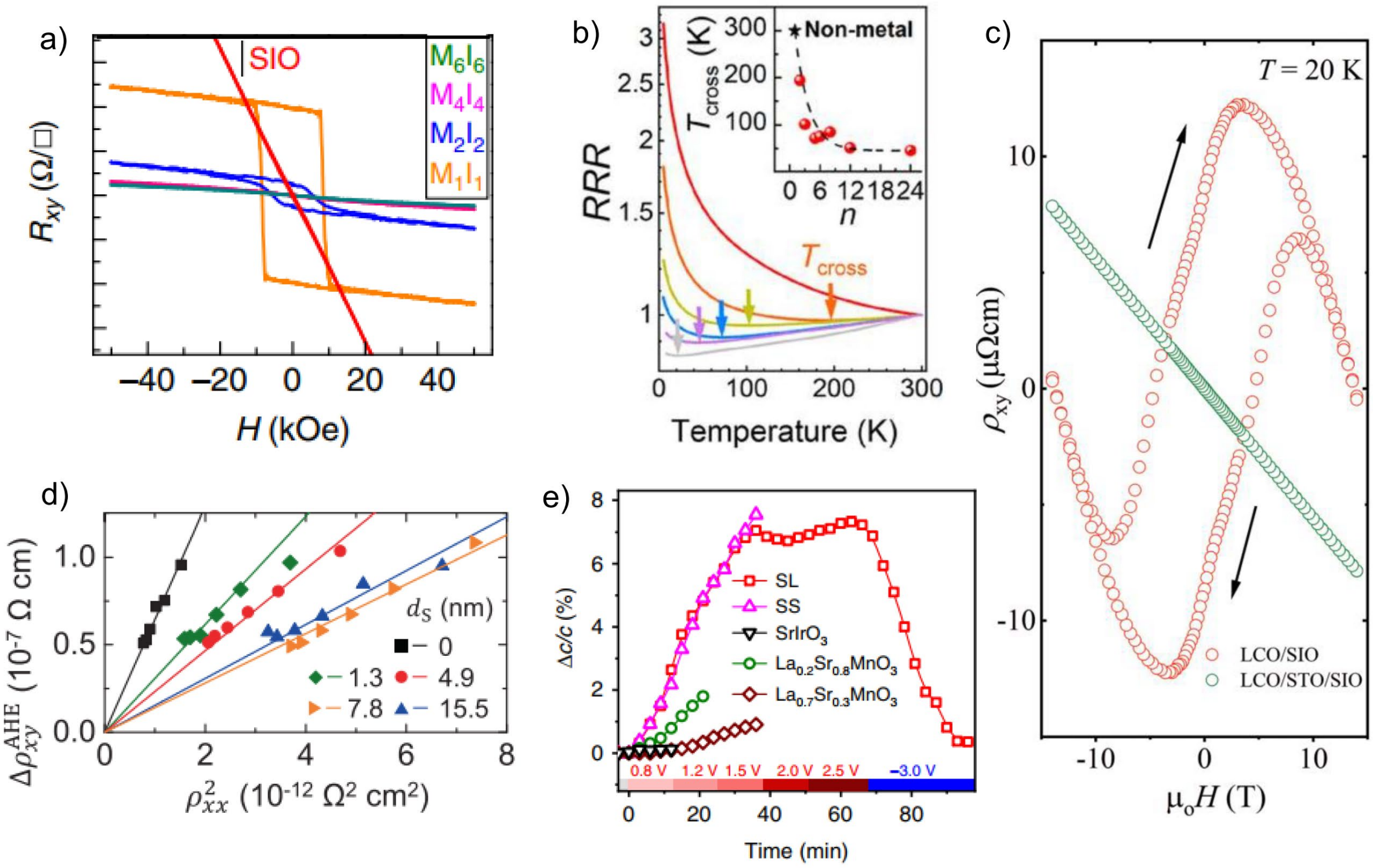

**Figure 18.** a) Anomalous Hall resistivity as a function of *c*-axis magnetic field of a series of $(SrIrO_3)$y/$(SrMnO_3)$x (MxIy in the notation) superlattices, reproduced with permission.[212] Copyright 2016, Springer Nature. b) Temperature dependent resistivity of $SrIrO_3/CaMnO_3$, reproduced with permission.[218] Copyright 2022, American Chemical Society. c) Field dependent anomalous Hall resistivity of $SrIrO_3/LaCoO_3$ and $SrIrO_3/SrTiO_3/LaCoO_3$ heterostructures, reproduced with permission.[213] Copyright 2022, Wiley-VCH. d) Relation between anomalous Hall resistivity and the square of longitudinal resistivity of $SrIrO_3/La_{0.7}Sr_{0.3}MnO_3$ heterostructures, reproduced with permission.[207] Copyright 2021, Springer Nature. e) Time evolution of the lattice structure of $SrIrO_3/La_{0.8}Sr_{0.2}MnO_3$ superlattice under ionic liquid gating, reproduced with permission.[211] Copyright 2022, Springer Nature.

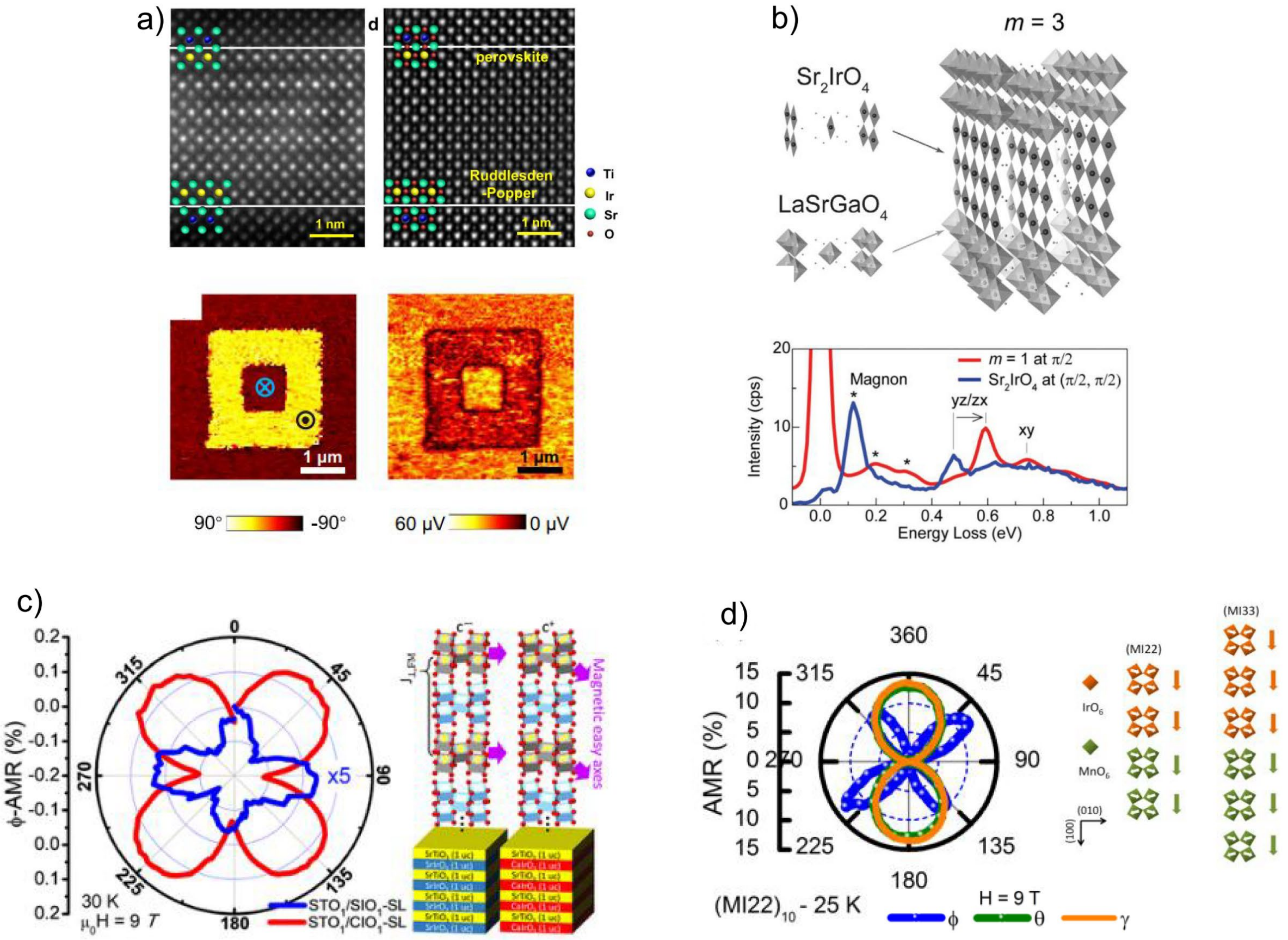

**Figure 19.** a) Atomically resolved High angle annular dark-field and integrated differential phase contrast images of $(Sr_2IrO_4)3/(SrTiO_3)6$ superlattice. The bottom panel shows the out-of-plane PFM images of the superlattice at 3.7 K, reproduced with permission.[268] Copyright 2021, Springer Nature. b) Schematic diagram of the *a*-axis-oriented $(Sr_2IrO_4)3/(LaSrGaO_4)5$ superlattice. The bottom panel displays resonant inelastic X-ray scattering spectra of the superlattice and $Sr_2IrO_4$, reproduced with permission.[269] Copyright 2016, Wiley-VCH. c) Anisotropic magnetoresistance effect and schematic of $(CaIrO_3)1/(SrTiO_3)1$, reproduced with permission.[272] Copyright 2020, American Chemical Society. d) Anisotropic magnetoresistance effect and magnetic structure of $CaIrO_3/CaMnO_3$, reproduced with permission.[273] Copyright 2022, American Physical Society.

## Acknowledgements

This work was supported by the Basic Science Research Programs through the National Research Foundation of Korea (NRF-2021R1A2C2011340 and 2022R1C1C2006723). Lin Hao acknowledges financial support from the Collaborative Innovation Program of Hefei Science Center, CAS (2022HSC-CIP005), and the HFIPS Director's Fund (2023YZGH01).